# Increased radiation events discovered at commercial aviation altitudes

**One Sentence Summary:** Energetic particles from the Van Allen radiation belts are the likely cause of enhanced radiation events experienced by aircraft at commercial aviation altitudes.

**Authors:** W. Kent Tobiska[1*], Alexa J. Halford[2], Steven K. Morley[3]

**Affiliations:**

[1]Space Environment Technologies, Space Weather Division, Pacific Palisades, California.

[2]The Aerospace Corporation, Chantilly, Virginia.

[3]Los Alamos National Laboratory, Los Alamos, New Mexico.

[*]Corresponding author: Email: ktobiska@spacenvironment.net.

**Abstract**: We show fifty-seven enhanced radiation level events taken from new measurements on commercial altitude (>9 km) aircraft that are analogous to planes flying through radiation clouds. More accurately, the plane is likely to be flying through a bremsstrahlung-origin γ-ray beam. Evidence points to the beam being produced at higher altitudes by incident relativistic electrons coming from the Van Allen radiation belts and that have been generated by electromagnetic ion cyclotron (EMIC) waves. The EMIC waves have been inferred by ground observatory, aircraft air, and satellite space observations as well as from modeling. We do not rule out other radiation sources associated with geomagnetic substorms and radiation belt coupling although these enhanced radiation events seem to frequently occur even during very minor geomagnetic disturbed conditions. These events show a dynamic and variable radiation environment at aircraft altitudes in a narrow magnetic latitude band (43° – 67° N, S; 2 < L-shell < 7). Observations are of dose rate enhancements that are statistically significant above the galactic cosmic ray (GCR) background. Measurements indicate that between 11–12 km during quiet geomagnetic conditions and at L-shell 4, the effective dose rate can be 32% higher than from GCRs alone. For the events themselves the mean effective dose rate is nearly double that for the background GCR level. The implication is that background exposure rates for North Atlantic (NAT), North Pacific (NoPAC), and the northern half of continental U.S. (CONUS) air traffic routes are typically higher than if one only considers GCRs. The net effect on aircraft crew and frequent flyers for these routes will be an increase in the monthly and annual exposures, which may have career-limiting health consequences.

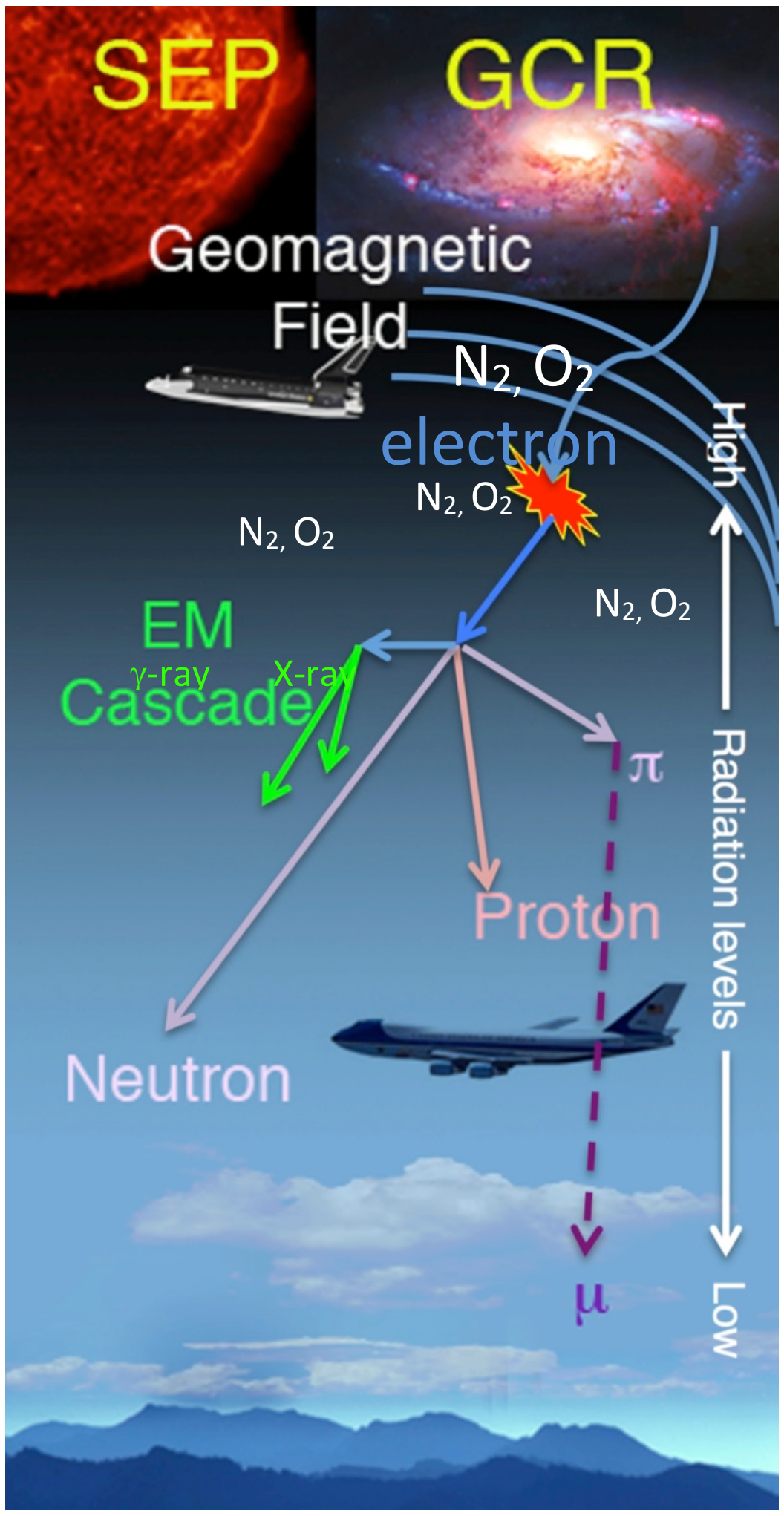

Fig. 1. The complex radiation conditions at and above commercial aviation altitudes from GCRs and SEPs (Tobiska et al., 2015).

**Background to atmospheric radiation**: Two major sources of radiation hazards have been known for decades, i.e., galactic cosmic rays (GCRs) and solar energetic particles (SEPs) (Figure 1). GCRs are produced outside the solar system in high-energy explosive events and consist mostly of energetic protons slowly modulated by the strength of the Sun's interplanetary magnetic field (IMF). SEPs come from either solar coronal mass ejections (CMEs) related to flaring events or from IMF shocks (Gopalswamy, 2004; Reames, 2013). In the latter case fast CMEs plow through a slower solar wind creating a shock front to produce energetic protons. In this paper, we describe a new, third radiation source likely originating from relativistic electrons that are associated with the Van Allen radiation belts. They create highly variable, dynamic mesoscale radiation events at aviation altitudes. Their manifestation is analogous to an aircraft flying through radiation "clouds," although physically the aircraft is flying through a $\gamma$-ray beam.

Independent of their source, charged particles such as protons ($p^+$), electrons ($e^-$), alpha particles ($\alpha$), and heavier ions including iron ($Fe^+$) arrive at the top of the atmosphere and create a primary radiation field. Depending upon their energies, which are a function of their mass and velocity squared, they are sorted according to the Earth's magnetic field standoff strength, or cutoff rigidity, and thus they can enter the Earth's atmosphere at different magnetic latitudes. Only higher energy particles can make it to low latitudes while even low energy particles can enter the atmosphere at higher latitudes. Because there are more of the lower energy particles (greater flux), the result is a higher radiation environment at higher magnetic latitudes.

Below the terrestrial atmosphere's mesopause near 85 km, the incident particles increasingly interact with neutral species, which are predominantly $N_2$ and $O_2$. The particle collisions with these target molecules create a spray of secondary and tertiary particles as well as photons. These include neutrons (n), $p^+$, $e^-$, $\alpha$, pions ($\pi$), muons ($\mu$), $\gamma$-rays and X-rays but each has lower energy than the primary particle. For example, *electron* impact on target atom nuclei and shell electrons can slow the incident electron, where a high-energy photon is released via bremsstrahlung; *photon* interaction with target atom nuclei and shell electrons can further eject electrons, photons, or positrons. These are two important processes creating secondary and tertiary radiation fields (Reitz *et al.,* 1993; Truscott *et al.,* 2004; Xu *et al.,* 2018). As primary particles are increasingly absorbed at lower altitudes, they compete in ionization with the increasing secondary and tertiary population being simultaneously produced. This results in a maximum ionization rate between 20 and 25 km (65,000–82,000 ft.) called the Regener–Pfotzer maximum (Regener and Pfotzer, 1935; Carlson and Watson, 2014). Below this altitude, down to the Earth's surface, the dose rate from all radiation then decreases because of particle and photon absorption in an increasingly thick atmosphere. All these secondary and tertiary particles are able to collide with an aircraft hull and its interior components, people, or fuel to further alter the radiation spectrum that is eventually felt by tissue (IARC, 2000; UNSCEAR, 2000) and avionics.

Atmospheric ionizing radiation is the primary source of human exposure to high linear energy transfer (LET) radiation at commercial aircraft altitudes (Wilson *et al.,* 1995; ECRP, 2004; Tobiska *et al.,* 2015). High LET radiation directly breaks up DNA, i.e., a process that can lead to adverse health effects including cancer. It has been noted that this can be an important cause for limiting the careers of aircrew (Jones *et al.,* 2005; Cannon, 2013).

**Scope of our study**: Based on an extensive study comparing *Automated Radiation Measurements for Aerospace Safety* (ARMAS) measurements (Tobiska *et al.,* 2018) with the NASA Langley Research Center (LaRC) *Nowcast of Atmospheric Ionizing Radiation for Aviation Safety* (NAIRAS) model (Mertens *et al.,* 2013), which extracted both GCR and suggested relativistic electron precipitation (REP) statistics, we wanted to better understand the nature of the postulated REP-source radiation measurements. Here we present a subset of observations of fifty-seven significant short-term events of dose rate enhancements occurring between 9 and 14 km altitude that cannot be attributed to GCRs or SEPs. In one case, we explore how energetic electrons from the outer Van Allen radiation belt can be precipitated by electromagnetic ion cyclotron (EMIC) waves (Tsurutani *et al.,* 2016) associated with particle injections from geomagnetic substorms. Using modeling and observations we show that EMIC waves may supply a population of relativistic electrons that subsequently become a source of radiation field enhancements, above the GCR background, and whose secondaries are observed by aircraft at commercial aviation altitudes. Although we have not identified all processes leading to this phenomenon we present four lines of evidence for an energetic electron radiation source using one example (Figure 2) that we consider typical.

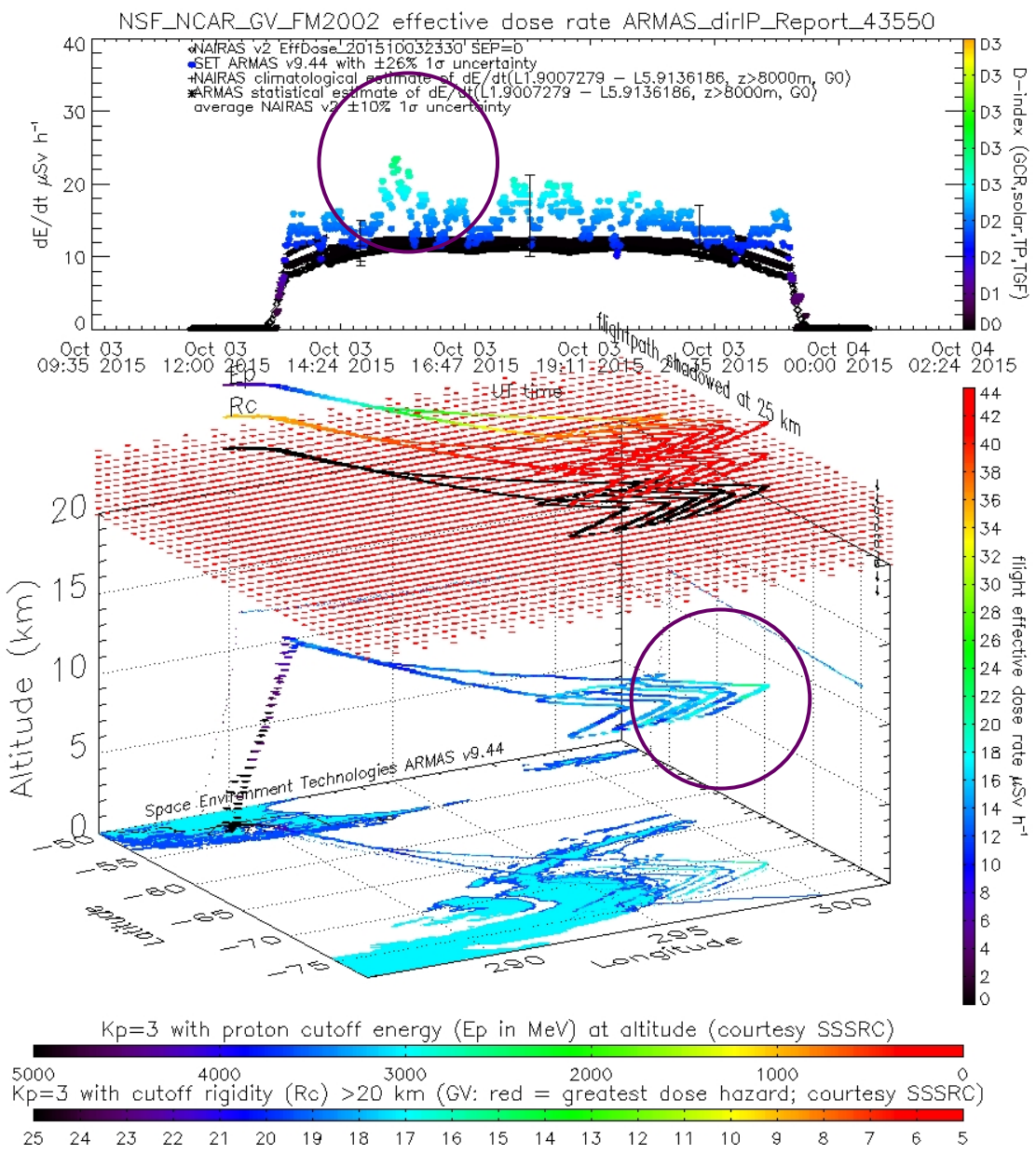

Fig. 2. The ARMAS 03 October 2015 G5 flight at 11.5 km and magnetic latitudes near –63° with doubled effective dose rate for ~1/2 hour; effective dose rate vs. time compared with NAIRAS (top panel black diamonds) and the flight geographical context with information related to proton cutoff energies and cutoff rigidities (bottom panel).

**First line of evidence (elevated dose rate measurements at commercial altitudes)**: The difficult task of continuous radiation environment monitoring, reporting, and modeling has not yet been achieved on regional or global scales. There are no existing routine dose measurements at aviation altitudes, either for tissue-relevant or avionics applications. Because of this, the ARMAS program began developing a calibrated, real-time, global monitoring, reporting, and modeling capability of the aviation radiation environment in 2012 (Tobiska *et al.,* 2016).

By early 2019 ARMAS has obtained real-time radiation measurements from the ground to 89 km for 599 flights consisting of 533655 one-minute measured absorbed dose (silicon) and derived effective dose rate records. The ARMAS monitoring system (Tobiska *et al.,* 2016; 2018) consists of two components: *i)* a flight instrument that measures the environment absorbed dose in silicon on an aircraft and *ii)* a real-time data stream from the aircraft to the ground, which is then processed to Level 4 effective dose rates for location and time. Measurements are made using the Teledyne micro dosimeter UDOS001 (μDos) in combination with a microprocessor, a GPS chip, an Iridium transceiver or a Bluetooth transmitter, and associated electronics. The μDos chip is sensitive to heavy ions ($Fe^{+}$), alphas, protons, neutrons, electrons, and γ-rays, especially above 1 MeV based on our extensive ground beam line testing (Tobiska *et al.,* 2016). All these components are mated to a printed circuit board and housed in a milled aluminum case. Once the absorbed dose (Si) is measured within the aircraft, it is relayed to the ground via Iridium satellite link or aircraft WiFi.

We use in this study only the science quality measured absorbed dose on the aircraft. Occasional data losses, e.g., no data from undelivered Iridium packets or cessation of operations from temperatures outside the range of –17° to +30° C are removed. Data corruptions in the form of anomalously high data values can occur and are produced by excessive electromagnetic interference (EMI) from nearby instruments on research aircraft or from nearby high-power military-grade radar. Those data are easily identifiable and not used in our analysis. Events that we consider as candidates for enhanced radiation due to REPs or other possible radiation belt particle precipitations all have a common feature, i.e., the data in each example starts near statistical and climatological GCR baseline values and then the dose rates are observed to rise significantly above and then later fall back to that baseline over ~1 hour (Figure 2, top panel).

Of particular interest as radiation event examples are the measurements in fifty-seven cases listed in Table 1. Aircraft making the measurements include the National Science Foundation (NSF) National Center for Atmospheric Research (NCAR) Gulfstream 5 (G-5), the National Oceanic and Atmospheric Administration (NOAA) G-4, the NASA Armstrong Flight Research Center (AFRC) DC-8 and G-3, the Federal Aviation Administration (FAA) William J. Hughes Technical Center (WJHTC) Bombardier Global 5000 (BG5) and commercial B-737 aircraft (B737). The data obtained were while NSF, NOAA, NASA, and FAA flights conducted science and engineering missions and where personnel from collaborating institutions flew on commercial flights. Commercial airport security is not an issue for ARMAS. For commercial flights the ARMAS Flight Modules (FMs) are typically screened by an X-ray machine upon entering the terminal waiting area and the Lithium-ion polymer battery power and safety features are compliant with aircraft safety standards; spec sheet documentation is included within the instrument carry-on case. The criteria used to select these events include altitudes above 9 km, level flight during the event, magnetic latitudes corresponding to inner and outer belt L-shells (magnetic latitudes between 43° and 67° for both hemispheres, i.e., L = 1.9 to 6.8), and a significant event rise (at least 40%) above then return to GCR background values as represented by modeled climatology for a flight. We did not observe these types of events at lower latitudes.

Table 1. ARMAS enhanced radiation events

| Event # and aircraft | Event time YYYY/MO/DY hh:mm:ss (UT) | Event LCT[8], duration (minutes)[9] | Event altitude (m) | Event dE/dt[10] | Event[11], flight[12] ratio to bkg | Event L, magnetic latitude[13] | Kp, Ap, NOAA G scale, Dst (nT) |
|---|---|---|---|---|---|---|---|
| **1. G-4[1]** | 2017/09/17 22:40:40 | 17:59, 25* | 14200 ±100 | 25 | 1.8, 1.4 | 2.2†, +47.7 | 4, 29, G0, –20 |
| **2. G-4[1]** | 2017/09/25 09:36:30 | 04:57, 27 | 14000 ±20 | 20 | 1.6, 1.5 | 2.2†, +47.9 | 4, 29, G0, –10 |
| **3. G-4[1]** | 2017/09/25 10:40:30 | 05:47, 34 | 14300 ±50 | 21 | 1.7, 1.5 | 2.2†, +47.2 | 4, 29, G0, –10 |
| **4. G-4[1]** | 2017/09/25 11:38:30 | 06:31, 38 | 14400 ±20 | 22 | 2.2, 1.5 | 1.9†, +43.0 | 4, 29, G0, –10 |
| **5. G-4[1]** | 2017/09/25 20:14:10 | 15:24, 26 | 13400 ±20 | 24 | 1.9, 1.3 | 2.5†, +50.3 | 4, 29, G0, –05 |
| **6. G-4[1]** | 2017/09/25 20:42:10 | 15:58, 24 | 13700 ±20 | 21 | 1.6, 1.3 | 2.6†, +51.3 | 4, 29, G0, –02 |
| **7. G-5[2]** | 2015/09/17 17:24:20 | 10:19, 15 | 14200 ±20 | 22 | 1.9, 1.2 | 2.2†, +47.8 | 3, 12, G0, –19 |
| **8. G-5[2]** | 2015/09/24 13:28:50 | 08:15, 60 | 11750 ±50 | 28 | 2.9, 1.5 | 2.6†, –51.9 | 2, 06, G0, –13 |
| **9. G-5[2]** | 2015/09/24 21:43:50 | 16:36, 38 | 11800 ±100 | 28 | 3.4, 1.5 | 2.4†, –49.3 | 2, 06, G0, –11 |
| **10. G-5[2]** | 2015/09/29 15:35:50 | 13:32, 99 | 11600 ±200 | 23 | 1.9, 1.4 | 6.0, –65.9 | 1, 02, G0, –01 |
| **11. G-5[2]** | 2015/09/29 16:30:50 | 14:50, 22 | 11400 ±50 | 23 | 1.8, 1.4 | 5.2, –63.9 | 1, 02, G0, –02 |
| **12. G-5[2]** | 2015/09/29 19:23:50 | 16:06, 33 | 11950 ±50 | 22 | 2.0, 1.4 | 3.3, –56.7 | 1, 02, G0, –02 |
| **13. G-5[2]** | 2015/10/03 15:29:10 | 11:38, 33 | 11550 ±50 | 26 | 2.2, 1.5 | 4.8, –62.7 | 3, 22, G0, –24 |
| **14. G-5[2]** | 2015/10/05 15:19:20 | 08:27, 33 | 11400 ±20 | 23 | 1.9, 1.3 | 5.0, –63.3 | 3, 19, G0, –27 |
| **15. G-5[2]** | 2015/10/10 16:05:10 | 11:58, 22 | 11450 ±50 | 25 | 2.1, 1.4 | 3.9, –59.4 | 3, 13, G1, –33 |
| **16. G-5[2]** | 2015/10/10 17:38:10 | 13:38, 21 | 11500 ±20 | 20 | 1.7, 1.4 | 3.5, –57.5 | 3, 13, G1, –33 |
| **17. G-5[2]** | 2015/10/10 19:47:10 | 15:29, 33 | 11550 ±50 | 22 | 1.9, 1.4 | 3.5, –57.5 | 3, 13, G1, –35 |
| **18. G-5[2]** | 2015/10/12 16:48:50 | 10:20, 38 | 11350 ±50 | 31 | 2.9, 1.4 | 6.2, –66.4 | 4, 23, G0, –28 |
| **19. DC-8[3]** | 2013/08/19 16:48:20 | 10:18, 56 | 9600 ±20 | 52 | 8.4, 2.2 | 2.3†, +48.4 | 1, 03, G0, –07 |
| **20. DC-8[3]** | 2013/08/28 00:24:10 | 17:43, 19 | 12600 ±300 | 19 | 2.0, 1.3 | 2.3†, +49.0 | 2, 09, G1, –44 |
| **21. DC-8[3]** | 2014/09/03 14:48:40 | 08:30, 21 | 11450 ±50 | 20 | 2.3, 1.6 | 2.3†, +49.2 | 3, 09, G0, +03 |
| **22. DC-8[3]** | 2014/09/03 15:52:40 | 08:54, 22 | 11500 ±20 | 20 | 2.5, 1.6 | 2.1†, +46.3 | 3, 09, G0, –01 |
| **23. DC-8[3]** | 2015/12/01 21:32:50 | 13:18, 25 | 11800 ±20 | 18 | 1.8, 1.2 | 2.8†, +53.1 | 2, 10, G0, –22 |
| **24. DC-8[3]** | 2015/12/02 00:19:50 | 16:13, 34 | 11800 ±20 | 19 | 1.8, 1.2 | 2.8†, +53.5 | 2, 10, G0, –18 |
| **25. DC-8[3]** | 2015/12/12 16:43:40 | 08:31, 36 | 11750 ±50 | 19 | 1.9, 1.2 | 2.6†, +52.0 | 2, 08, G0, –06 |
| **26. DC-8[3]** | 2015/12/12 17:26:40 | 09:19, 20 | 11700 ±100 | 19 | 1.8, 1.2 | 2.7†, +52.6 | 2, 08, G0, –05 |
| **27. DC-8[3]** | 2015/12/12 19:47:40 | 11:45, 28 | 11700 ±20 | 19 | 1.8, 1.2 | 2.8†, +52.9 | 2, 08, G0, –04 |

| | | | | | | | | |
|---|---|---|---|---|---|---|---|---|
| **28.** | **DC-8[3]** | 2017/06/28 14:53:40 | 09:20, 31 | 11350 ±50 | 25 | 2.7, 1.3 | 2.2†, +47.4 | 3, 14, G0, –00 |
| **29.** | **DC-8[4]** | 2016/06/10 10:29:10 | 22:04, 13 | 10700 ±50 | 25 | 2.5, 1.9 | 2.4†, –49.6 | 2, 11, G0, +07 |
| **30.** | **DC-8[4]** | 2016/06/10 11:01:10 | 22:14, 24 | 10700 ±50 | 30 | 3.0, 1.9 | 2.4†, –49.6 | 2, 11, G0, +10 |
| **31.** | **DC-8[4]** | 2016/06/10 11:40:10 | 22:32, 20 | 10700 ±50 | 25 | 2.6, 1.9 | 2.4†, –49.8 | 2, 11, G0, +13 |
| **32.** | **DC-8[4]** | 2016/06/10 12:15:10 | 23:07, 21 | 10700 ±50 | 21 | 2.4, 1.9 | 2.4†, –49.8 | 2, 11, G0, +15 |
| **33.** | **DC-8[4]** | 2016/10/14 23:26:10 | 18:59, 20* | 11300 ±50 | 20 | 2.4, 1.9 | 2.1†, –46.6 | 2, 11, G0, –12 |
| **34.** | **DC-8[4]** | 2016/10/20 23:17:10 | 18:26, 20 | 10500 ±50 | 18 | 2.0, 1.2 | 2.5†, –50.8 | 2, 11, G0, +03 |
| **35.** | **DC-8[4]** | 2016/10/31 15:34:10 | 09:44, 35 | 9700 ±50 | 26 | 2.9, 1.4 | 3.4, –57.3 | 2, 11, G0, –24 |
| **36.** | **DC-8[4]** | 2016/10/31 20:17:10 | 11:58, 24 | 10900 ±50 | 24 | 2.0, 1.4 | 6.8, –67.4 | 2, 11, G0, –16 |
| **37.** | **DC-8[4]** | 2016/11/02 20:32:10 | 13:26, 27 | 10850 ±25 | 24 | 2.0, 1.2 | 5.8, –65.4 | 2, 11, G0, –20 |
| **38.** | **DC-8[4]** | 2016/11/04 15:35:10 | 10:17, 23 | 9050 ±50 | 17 | 1.8, 1.6 | 4.4, –61.5 | 2, 11, G0, –12 |
| **39.** | **DC-8[4]** | 2016/11/07 21:02:10 | 14:01, 31 | 11100 ±100 | 21 | 1.4, 1.3 | 4.9, –63.1 | 2, 11, G0, +01 |
| **40.** | **DC-8[4]** | 2016/11/07 23:33:10 | 18:24, 24 | 11300 ±50 | 20 | 2.3, 1.3 | 2.1†, –46.7 | 2, 11, G0, –00 |
| **41.** | **DC-8[4]** | 2016/11/09 14:38:10 | 09:43, 20 | 9400 ±20 | 18 | 1.9, 1.3 | 2.9†, –53.9 | 2, 11, G0, –03 |
| **42.** | **B737[5]** | 2016/04/10 19:08:10 | 12:22, 20 | 11300 ±50 | 16 | 1.5, 1.4 | 2.4†, +49.8 | 2, 11, G0, +02 |
| **43.** | **B737[5]** | 2016/04/25 14:24:10 | 05:19, 20 | 11900 ±50 | 18 | 2.6, 1.3 | 1.9†, +44.1 | 2, 11, G0, +04 |
| **44.** | **B737[5]** | 2016/08/01 20:34:10 | 13:43, 33 | 10100 ±50 | 20 | 2.6, 1.5 | 2.0†, +45.0 | 2, 11, G0, +04 |
| **45.** | **B737[5]** | 2016/11/12 23:44:10 | 13:15, 23 | 10350 ±20 | 40 | 3.3, 1.1 | 4.7, +62.5 | 2, 11, G0, –19 |
| **46.** | **B737[5]** | 2016/11/13 00:38:10 | 12:56, 26 | 10350 ±20 | 54 | 4.3, 1.1 | 4.3, +61.1 | 2, 11, G0, –17 |
| **47.** | **B737[5]** | 2017/02/16 23:52:10 | 15:00, 22 | 11000 ±50 | 24 | 2.0, 1.0 | 4.2, +60.9 | 2, 11, G0, +06 |
| **48.** | **B737[5]** | 2017/02/17 00:30:10 | 15:10, 22 | 11600 ±50 | 20 | 1.4, 1.0 | 5.1, +63.8 | 2, 11, G0, +06 |
| **49.** | **B737[5]** | 2017/02/18 21:28:10 | 16:15, 21 | 10700 ±50 | 17 | 1.7, 1.2 | 2.8†, +53.3 | 2, 11, G0, +01 |
| **50.** | **B737[5]** | 2017/02/20 03:29:10 | 20:40, 20 | 10350 ±20 | 20 | 1.8, 1.1 | 3.1, +55.3 | 2, 11, G0, –15 |
| **51.** | **B737[5]** | 2017/05/06 00:15:10 | 14:30, 40 | 10350 ±20 | 26 | 2.4, 2.9 | 3.1, +55.3 | 2, 11, G0, +15 |
| **52.** | **B737[5]** | 2017/06/10 22:40:10 | 13:58, 24 | 10350 ±20 | 18 | 1.8, 1.1 | 3.3, +56.8 | 2, 11, G0, +06 |
| **53.** | **B737[5]** | 2017/06/11 02:33:10 | 14:21, 25 | 11000 ±50 | 18 | 1.8, 1.1 | 2.6†, +51.3 | 2, 11, G0, +13 |
| **54.** | **G-3[6]** | 2017/08/21 17:00:10 | 08:41, 20 | 10670 ±20 | 17 | 1.9, 0.9 | 2.4†, +49.6 | 2, 11, G0, –04 |
| **55.** | **BG5[7]** | 2017/03/30 18:06:10 | 13:05, 25* | 12500 ±20 | 20 | 1.9, 1.7 | 2.2†, +47.7 | 2, 11, G0, –23 |
| **56.** | **BG5[7]** | 2017/03/30 18:38:10 | 13:55, 22 | 12500 ±20 | 20 | 2.0, 1.7 | 2.2†, +47.6 | 2, 11, G0, –25 |
| **57.** | **BG5[7]** | 2017/03/30 20:37:10 | 15:37, 22 | 12500 ±20 | 20 | 1.9, 1.7 | 2.2†, +47.6 | 2, 11, G0, –24 |

[1]The ARMAS Flight Module 2 (FM2B) on the NOAA Gulfstream 4 (G-4).

[2]The ARMAS Flight Module 2 (FM2A) on the NSF/NCAR Gulfstream 5 (G-5).

[3]The ARMAS FM1 on the NASA/AFRC DC-8 (DC-8).

[4]The ARMAS FM5A on the NASA/AFRC DC-8 (DC-8).

[5]The ARMAS FM5B on a commercial (B737) aircraft.

[6]The ARMAS FM5C on the NASA/AFRC Gulfstream 3 (G-3).

[7]The ARMAS FM5B on the FAA/WJHTC Bombardier Global 5000 (BG5).

[8]Local Clock Time (LCT).

[9]Duration includes both physical location of the radiation region plus flight through some part of it.

[10]The effective dose rate, dE/dt, in μSv $h^{-1}$ where ARMAS RMS 1-σ uncertainty is ±24% using ARMAS v9.44 algorithm.

[11]The event maximum effective dose rate ratioed to NAIRAS climatological background; mean = 2.25 for all events.

[12]The total flight effective dose rate ratioed to NAIRAS climatological background; mean = 1.43 for all flights.

[13]The L-shell and magnetic latitude at the minute of event maximum.

†This event may be from unknown primaries and processes since it coincides with an inner belt L-shell.

*Flight takeoff/landing coincided with the start/end of the observed event.

NOTE: gray highlighted event 13 is discussed in detail in the text.

The NAIRAS model produces data-driven, physics-based climatology of time-averaged global radiation conditions. It covers the entire domain of interest using physics-based modeling. It predicts dosimetric quantities from the surface of the Earth, through the atmosphere, and into LEO from both GCRs and SEPs. It includes the response of the geomagnetic field to interplanetary dynamical processes and subsequent influences on atmospheric dose. It uses coupled physics-based models to transport cosmic radiation through three distinct domains: the heliosphere, Earth's magnetosphere, and neutral atmosphere. The physics-based models are driven by real-time measurements to specify boundary conditions on the cosmic and solar radiation at the interfaces between the distinct domains or to characterize a domain's internal properties through which radiation propagates. Since SEPs are quite rare, NAIRAS provides an excellent data-driven climatological estimate of the GCR-induced radiation environment.

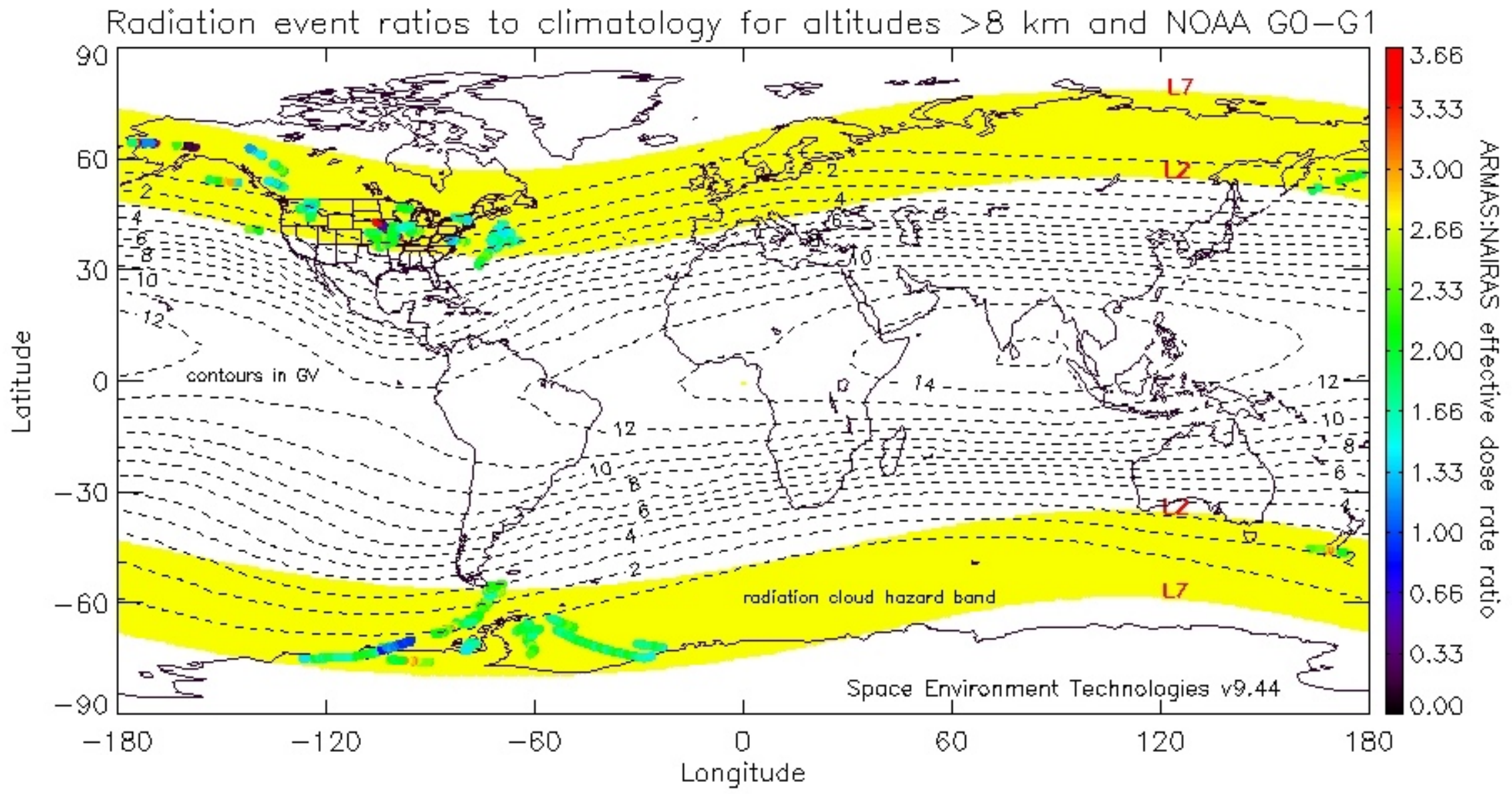

Fig. 3. One-minute ratios of ARMAS (weather) to NAIRAS (GCR climatology) for all radiation events identified in Table 1, for all altitudes above 8 km, and for geomagnetic conditions of G0 and G1 on the NOAA G-scale. Cutoff rigidities are in units of GV (dashed lines) (Shea and Smart, 2012).

Figure 3 shows the ratio of ARMAS (weather) to NAIRAS (GCR climatology) effective dose rates throughout each minute of every flight listed in Table 1. These ratios compare ARMAS to NAIRAS for each altitude level above 8 km and for geomagnetic conditions of G0 and G1 on the NOAA G-scale. Most ratios are above 1.0 and the mean of 1537 one-minute ratioed time records for all events is 1.91. In other words, there is nearly a factor of 2 increase in dose rate (green values) while flying through a radiation event compared with flying at the same altitude and under the same geomagnetic conditions under only GCR background conditions. In Figure 3, the data are also plotted in comparison with the global cutoff rigidity (Rc) contours (dashed lines) in units of GV (Shea and Smart, 2012). It is apparent that the events occur where Rc values are less than 4 GV, with most events occurring for ≤2 GV in both hemispheres. There effectively exists a defined magnetic latitude band within which these events occur. We have illustrated the L-shell (McIlwain, 1961; Tascione, 1994) range of 2–7, labeled "L2" and "L7," in yellow as notional bands in both hemispheres that we believe best fit the global locations where these events can be experienced. L-shells from ~3–7 are the regions where outer radiation belt charged particles map into the upper atmosphere while L-shells from ~1.3–2 are the regions where inner radiation belt charged particles map into the upper atmosphere. This L-shell band is consistent with Katsiyannis *et al.* (2018) who observed low Earth orbit relativistic electrons from the PROBA-V EPT and LYRA instruments and Dachev *et al.* (2017) as well as Dachev (2017) who have called these precipitation bands observed by the Liulin instrument on the ISS.

As an example of our data, we illustrate one typical flight (Figure 2, top panel) on 03 October 2015 at 15–16 UT where the event, circled in the figure and with gray highlight (#13) in Table 1, had an effective dose rate that rose and declined by a factor of two above GCR background levels within 33 minutes. The entire flight dataset was collected for 13.1 h between 03 October at 11:30 UT and 04 October at 00:34 UT. The event occurred for a fraction of the total flight. Uncertainties in the ARMAS data are ±24% and are described by Tobiska *et al.* (2016).

We note that for the range of cutoff rigidities from Chile to Antarctica (Rc = 3.89 to 0.24, respectively; Figure 4) the mean radiation quality factor (Q, relevant to the dose rate in tissue) was 2.03 with a Rc dependency (Tobiska *et al.,* 2016). The total flight effective dose was 157 μSv, i.e., the exposure of approximately one and a half chest X-rays. For this event starting at 15:07 UT (–60.1° magnetic latitude) the peak dose rate (top panel colored dots in Figure 2) occurred at 15:29 UT (11:38 Local Clock Time) when the aircraft was flying at an L-shell of 4.8 corresponding to –62.7° magnetic latitude. Background dose rates were once again reached at 15:45 UT (–64.6° magnetic latitude). The background levels were very close to the values of NAIRAS shown in the Figure 2 top panel (black "+" NAIRAS climatology, black "◇" NAIRAS values for date/time/location, black "✳" ARMAS statistical mean values for locations and geomagnetic conditions). These three background estimates lie nearly on top of one another and represent a consensus for the GCR contribution to the radiation environment. From the plane's perspective, which was flying at a constant altitude in this mesoscale region, the increased dose rate for 33 minutes was analogous to flying through a radiation cloud.

In previous work (Tobiska *et al.,* 2016; 2018) it was observed that the dose rate enhancement was not SEP-related since no solar event had occurred within the previous few days. In addition, changes in GCRs do not normally produce this magnitude of short-term and localized variability along a flight path. There were no large tropospheric storms or hurricanes in the region that might produce terrestrial γ-ray flashes (TGFs) or sprites (Chilingarian *et al.,* 2015) as a radiation source. Mertens *et al.* (2010) show that high geomagnetic activity can affect the aviation radiation environment. Tobiska *et al.* (2016) identified a minor Dst disturbance during the event and suggested that a geomagnetic pathway might create a dose rate peak. They speculated that energetic electron precipitation, induced by EMIC waves from the outer radiation belt, may be a cause for the +5σ observations above background. We explore below that event in greater detail.

In summary, there was essentially no change in background GCRs during each of the fifty-seven events for their short durations and there were no SEPs at the time of each event. From Table 1, the events demonstrate dynamic variability above the background GCR level over short time scales and across mesoscale regions. In the discussion below we conclude that these events must be connected to magnetospheric processes, which alter the radiation belt charged particle fluxes, i.e., the likely primary source behind the secondary radiation we are observing.

**Second line of evidence (space weather effects)**: Substorm injections can provide a source of MeV electrons in the outer radiation belt and are associated with high speed streams (HSS) of particles from the Sun (Tsurutani *et al.,* 2006; Hajra *et al.,* 2015a, 2015b). Direct injection of MeV particles by substorms has also been reported (Ingraham *et al.,* 2001). Exploring the 03 October 2015 example, we observed on 30 September 2015 that the Earth crossed the heliospheric current sheet and experienced a change in the polarity of the IMF. The Earth then encountered a long period over 5 days of disordered, slow solar wind produced by solar streamers and pseudo-streamers. For these rather typical interplanetary conditions, the 03 October 2015 magnetospheric response was moderate, as reflected by geomagnetic activity on this day peaking at Kp = 3+. At the time of the event the Dst index was –24 nT. No solar energetic proton events were reported during this period, and no enhancements were detected in either the NOAA GOES >10 MeV proton flux measurements or the Global Positioning System proton data. The GOES-13 0.6 MeV electrons were rising and the >1 MeV proton flux was elevated following a rapid enhancement during the 03 October observations (Figure 5). We interpret the >1 MeV protons showing a dispersed substorm particle injection, while the electrons show an enhanced flux following the injection. During geomagnetic substorms, convection strengthens very quickly across the entire magnetosphere (Miyashita *et al.,* 2008) and the rapid enhancement in GOES-13 1 MeV protons is consistent with their injection to at least geosynchronous orbit.

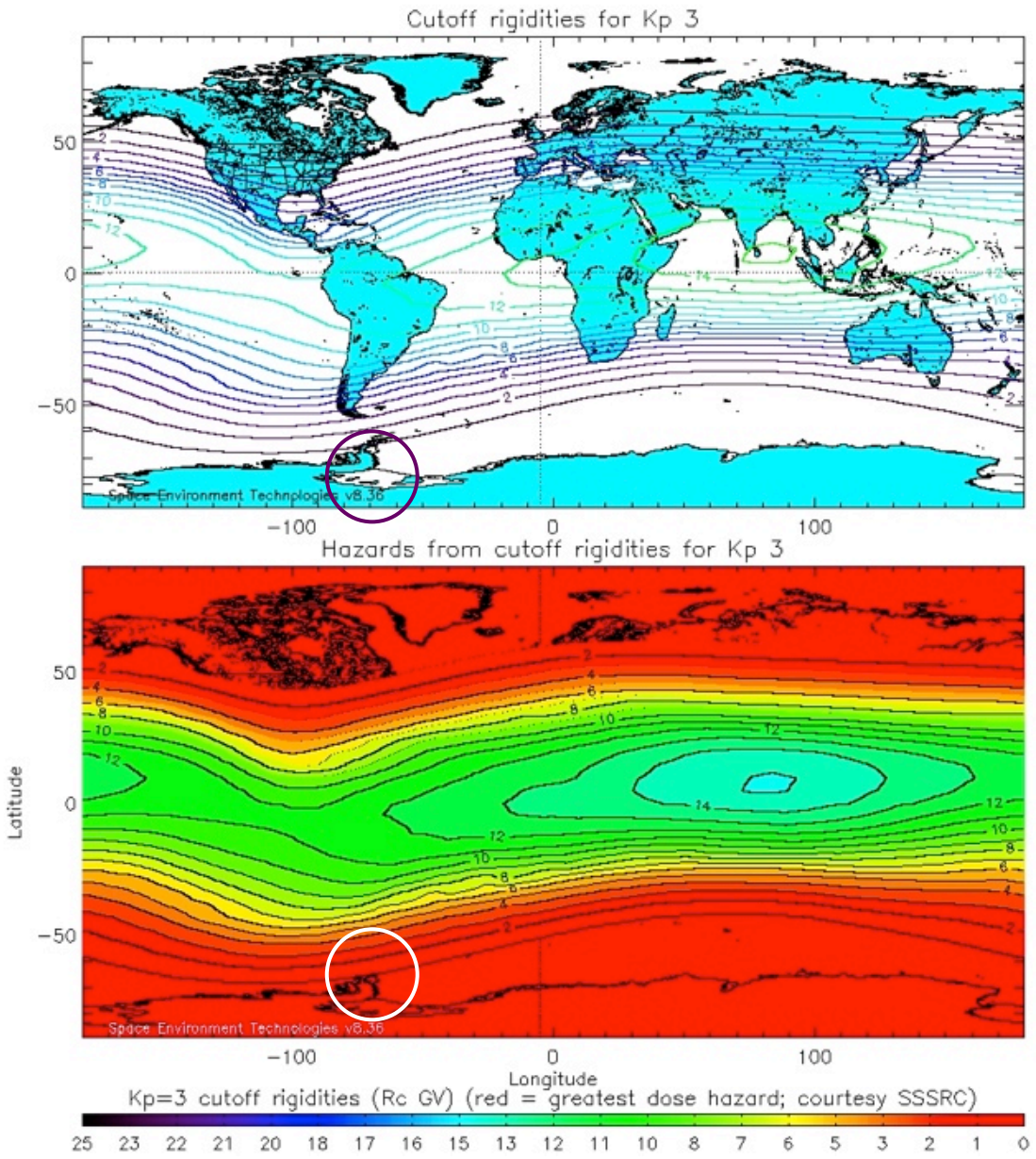

Fig. 4. The 03 October 2015 G5 flight vs. magnetic field cutoff rigidities (top panel) and hazard level (bottom panel) where red is the greatest hazard and aqua is the least (Shea and Smart, 2012). Circles identify the enhanced event location on 03 October 2015.

Substorm particle injections contribute significantly to radiation belt disturbances (Forsyth *et al.,* 2016; Tang *et al.,* 2016; Tsurutani *et al.,* 2016) and are conducive to EMIC growth and subsequent relativistic electron precipitation (Blum *et al.,* 2015). Both the outer and inner radiation belts were disturbed during this injection event, despite the moderate geomagnetic indices. This was evidenced by the NASA Radiation Belt Storm Probes (RBSP, now called Van Allen Probes, VAP) spacecraft A Relativistic Electron Proton Telescope (REPT) instrument that observed the >2.3 MeV electron population with fluxes of $1\times10^5$ and $3\times10^4$ electrons $cm^{-2}$ $s^{-1}$ $sr^{-1}$ $MeV^{-1}$ at L = 4.1 and 4.8, respectively (Figure 6). We exponentially fit the electron energy spectrum vs. flux, *j*, in equation 1 for the REPT measurements during the aircraft time frame. The fit over low pitch angles made the estimate more representative of fluxes near the loss cone. The exponential fit is:

$$j = A\ exp(-E/E_0) \qquad (1)$$

where $A = 6\times10^6$ and $E_0 = 0.46$; the derived fluxes of 2.3–5 MeV electrons at L = 4.1 and L = 4.8 (Figure 7) show a fit near 5 MeV of $1\times10^2$ electrons $cm^{-2}$ $s^{-1}$ $sr^{-1}$ $MeV^{-1}$, i.e., consistent with the REPT-measured spin-averaged fluxes at 5 MeV (Figure 8). We return to this fitted spectrum when considering the fourth line of evidence.

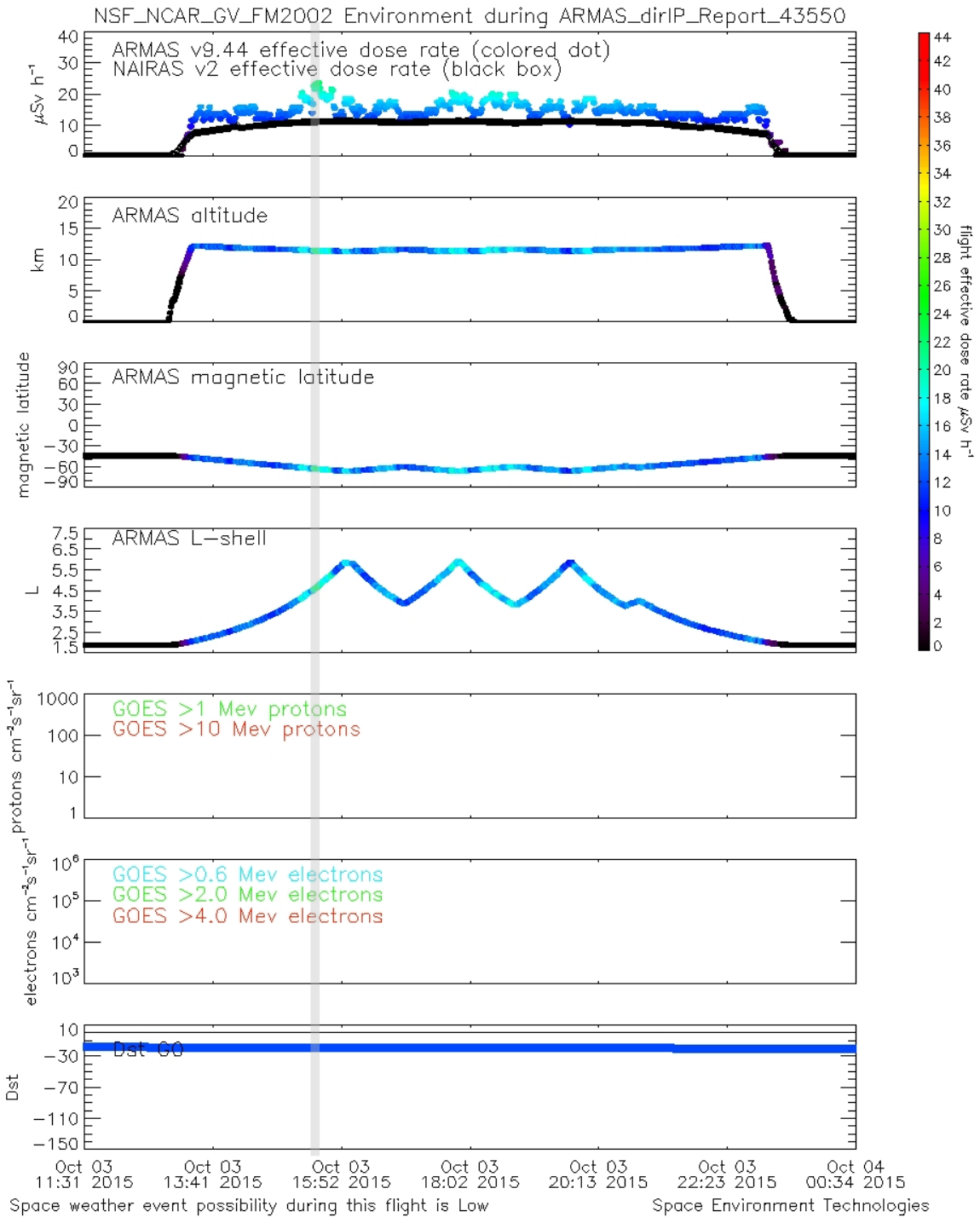

Fig. 5. Aircraft peak dose rate (top panel and vertical gray line) on October 03, 2015 (15 UT) during a substorm injection; altitude (2nd panel from top); magnetic latitude (3rd panel from top); L-shell (4th panel from top); GOES-13 >1 (green) and >10 (red – none) MeV protons (3rd panel from bottom); GOES-13 >0.6 (aqua), >2 (green – none), and >4 (red – none) MeV electrons (2nd panel from bottom) showing the rising particle injection; and Dst (bottom panel).

At L-shells above 4 in Figure 6 we see electron depletion between 02–03 October compared to refilling of this region starting 04 October. Thus, we infer that electron losses were similar to a loss event on 17 January 2013 described by Shprits *et al.* (2016). These losses were occurring before, during, and after the time of the aircraft measurements (vertical gray line). These losses are consistent with mechanisms of energetic electron dayside drift, radial diffusion to lower L-shells with a resultant energy increase, and atmospheric precipitation due to larger loss cone pitch angles. The spin-averaged electron fluxes are shown as energy vs. time (Figure 8, 2nd and 3rd panels). Energies extend from 2 to 20 MeV with the latter fluxes of $1\times10^3$ to $1\times10^4$ electrons $cm^{-2}$ $s^{-1}$ $sr^{-1}$ $MeV^{-1}$. This population is consistent with radial diffusion to higher energy values and lower L-shells.

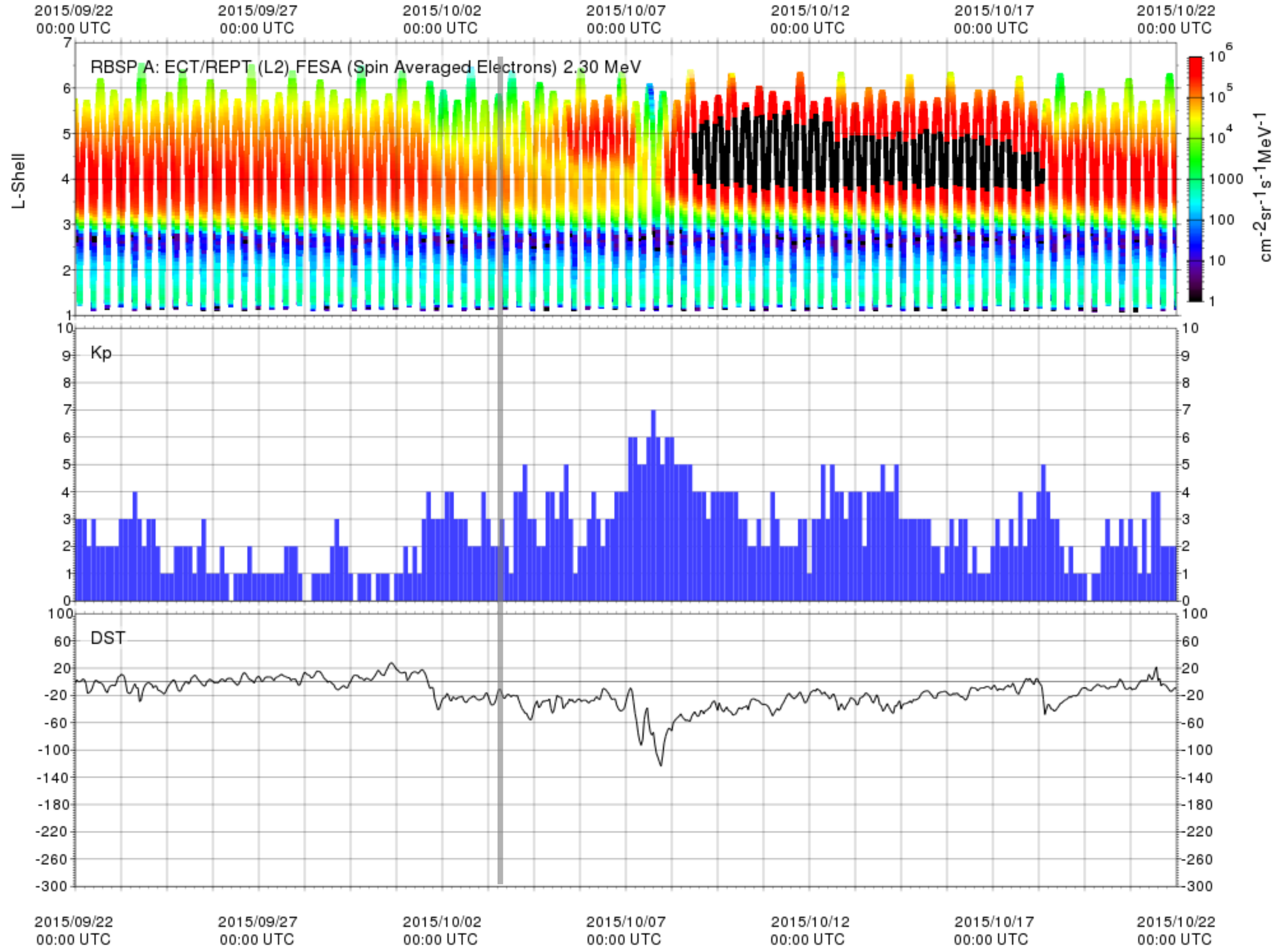

Fig. 6. RBSP A >2.3 MeV electrons vs. L-shell (top panel), Kp (middle panel), and Dst (bottom panel) from 22 September through 22 October 2015. The 02–03 October period contains a depletion of electrons due to particle injection from substorm main phase and radial diffusion with the peak aircraft event shown by the vertical gray line.

While RBSP A and B were in the midnight sector during ARMAS observations we see evidence for EMIC waves when VAP B passed noon prior to the flight in Figure 8. We conclude that >2 MeV relativistic electrons (Figure 8) indicate a disturbed electron population during the aviation observations.

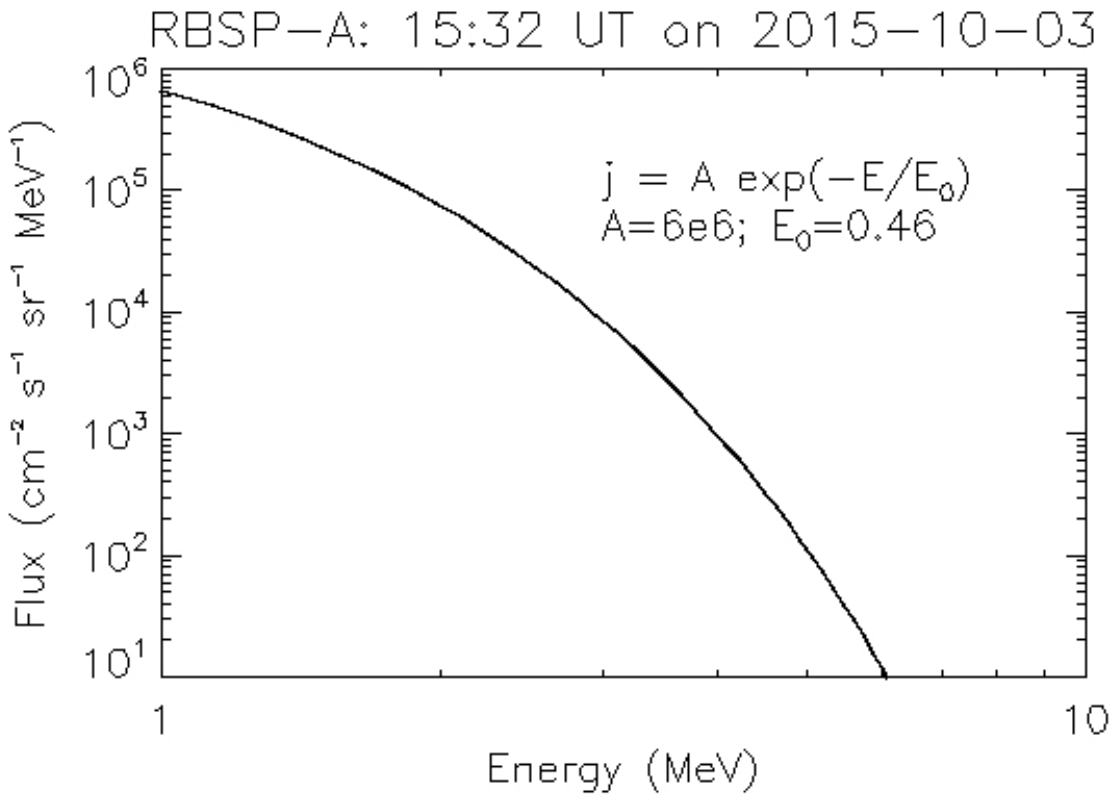

Fig. 7. RBSP A REPT exponential fit to injected electron energies vs. flux on 03 October 2015 at 15 UT; fit parameters are shown.

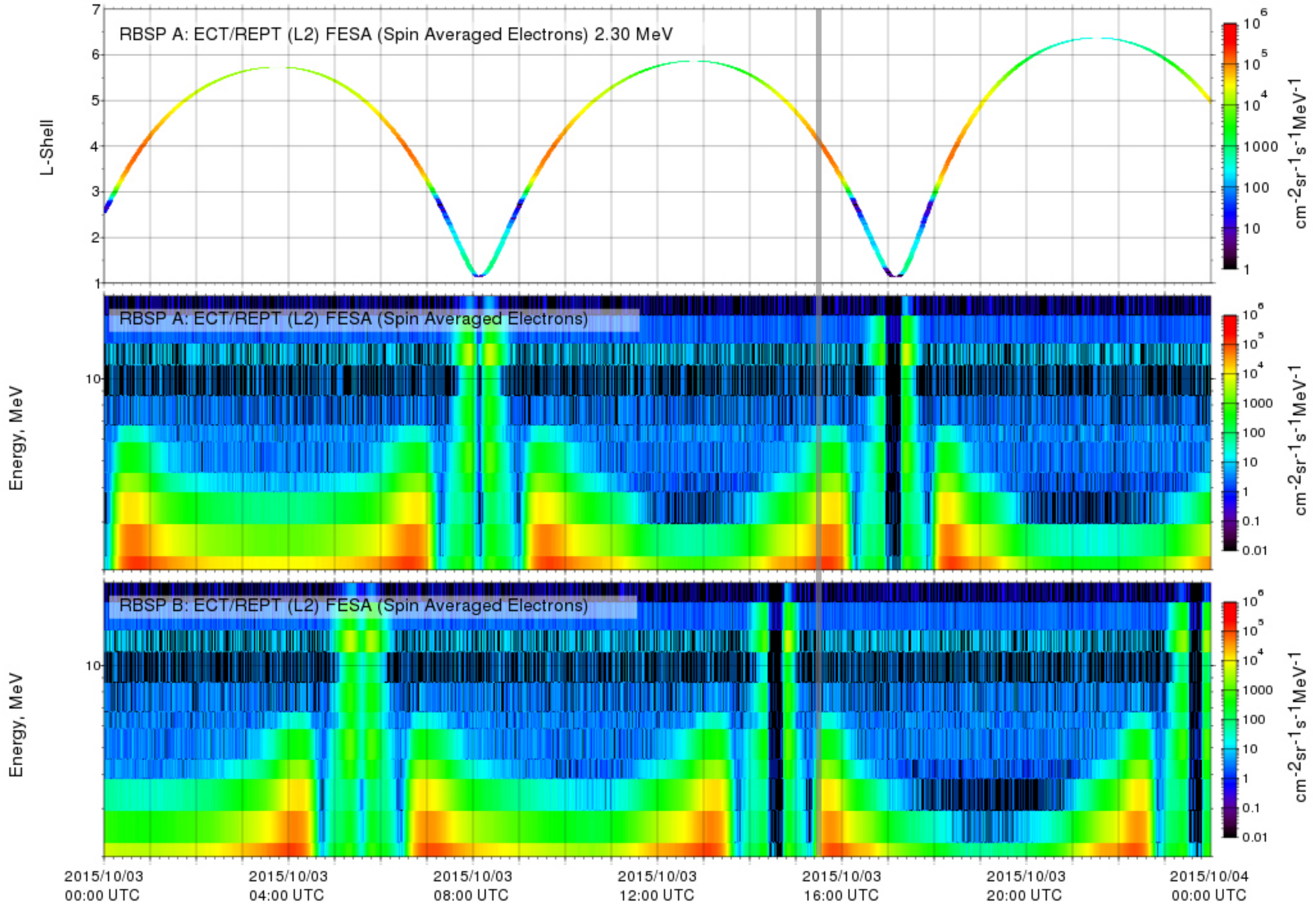

Fig. 8. RBSP A >2.3 MeV electron densities with L vs. time (top panel); A and B REPT spin-averaged electron densities (2nd and 3rd panels) with 10–20 MeV relativistic electrons from radial diffusion near L 2. The vertical gray line shows the peak of the aircraft radiation event.

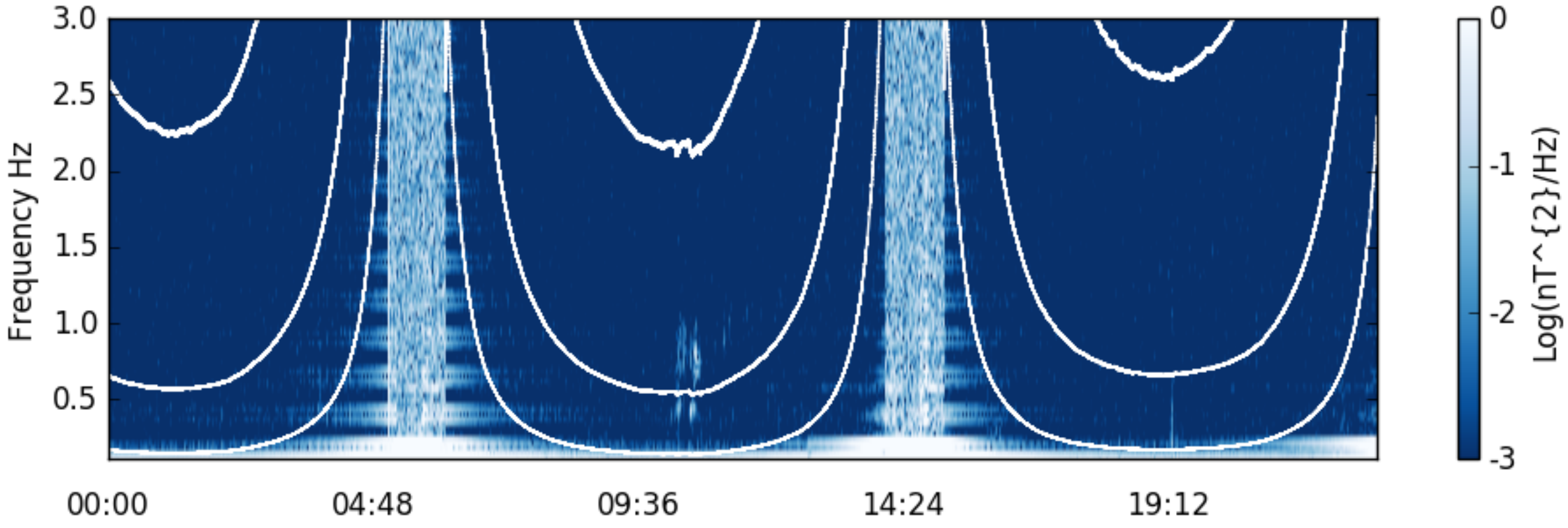

Fig. 9. EMIC waves in RBSP A orbit prior to the ARMAS 03 October 2015 observations.

**Third line of evidence (identification of EMIC waves as the mechanism for transporting electrons)**: EMIC waves can create populations of electrons that are able to precipitate into the atmosphere. Gaines *et al.* (1995) found from UARS HEP measurements that there were 2 orders of magnitude increase in >1 MeV electrons at L = 4 during geomagnetic storm periods followed by significant increases in relativistic electrons in the inner radiation belt (2 ≤ L ≤ 3) a day later. High fluxes of relativistic electrons existed at least a week after a major geomagnetic storm, even while Dst was –30 to –20 nT, and they concluded that ion pair production below 60 km can easily be an order of magnitude more than during quiet geomagnetic conditions. Clilverd *et al.* (2007) were further able to demonstrate dayside relativistic electron precipitation (REP) from EMIC waves, originating in the 4 ≤ L ≤ 5 outer radiation belt that were associated with a CME-magnetosphere coupled event. Xu *et al.* (2018) studied monoenergetic beams of 0.1–10 MeV precipitating electrons into the atmosphere to understand the process of bremsstrahlung radiation and its resultant ionization production and atmospheric effects, including pitch angle dependence in the ionization rate profile. In our study, we infer that higher fluxes of relativistic electrons were present in the radiation belt during this period of time. This is because the mechanisms for increasing energy to >2 MeV are a by-product of both EMIC wave and chorus wave local acceleration. Figure 9 shows EMIC waves from RBSP A during the orbit prior to the 03 October event with frequencies in the range of 0.3–1.0 Hz.

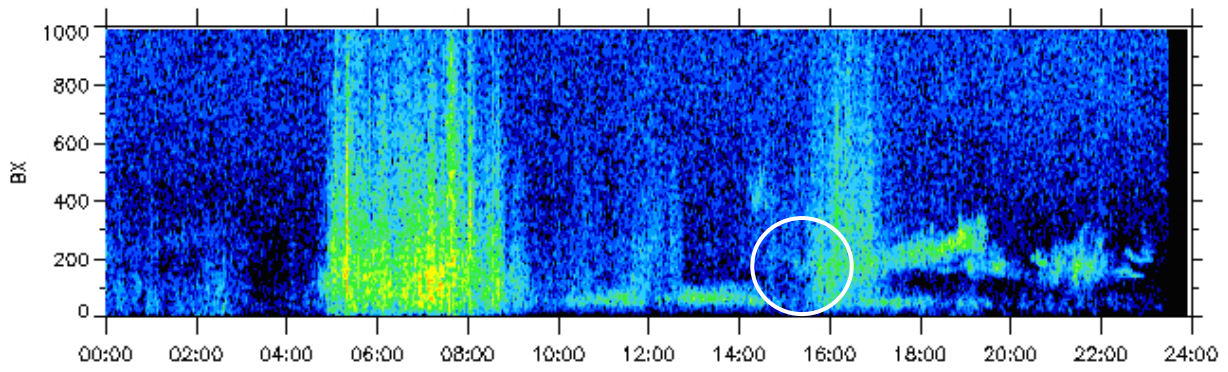

Fig. 10. ULF EMIC waves at 0.1 Hz (100 mHz) on 03 October 2015 at Halley Bay, Antarctica (L=4.3, circled area).

EMIC waves can be ducted through the ionosphere before reaching the ground (Kim *et al.,* 2011), thus allowing for their observations by ground-based magnetometers far from the field lines where they originate (Rodger *et al.,* 2007). Although this means we have lost the information about where the waves originated, we do know that waves were present in the magnetosphere at this time. Observational evidence of EMIC waves during the 03 October event comes from ground stations that observe ULF ionospheric frequencies, especially around 0.1 Hz (Cliverd *et al.,* 2009) in the horizontal (X, Y) components. For example, the magnetometers at Halley Bay, Antarctica (-64.2/-40.6 latitude/longitude, L = 4.3) showed EMIC waves near the 03 October event (Figure 10, circled area).

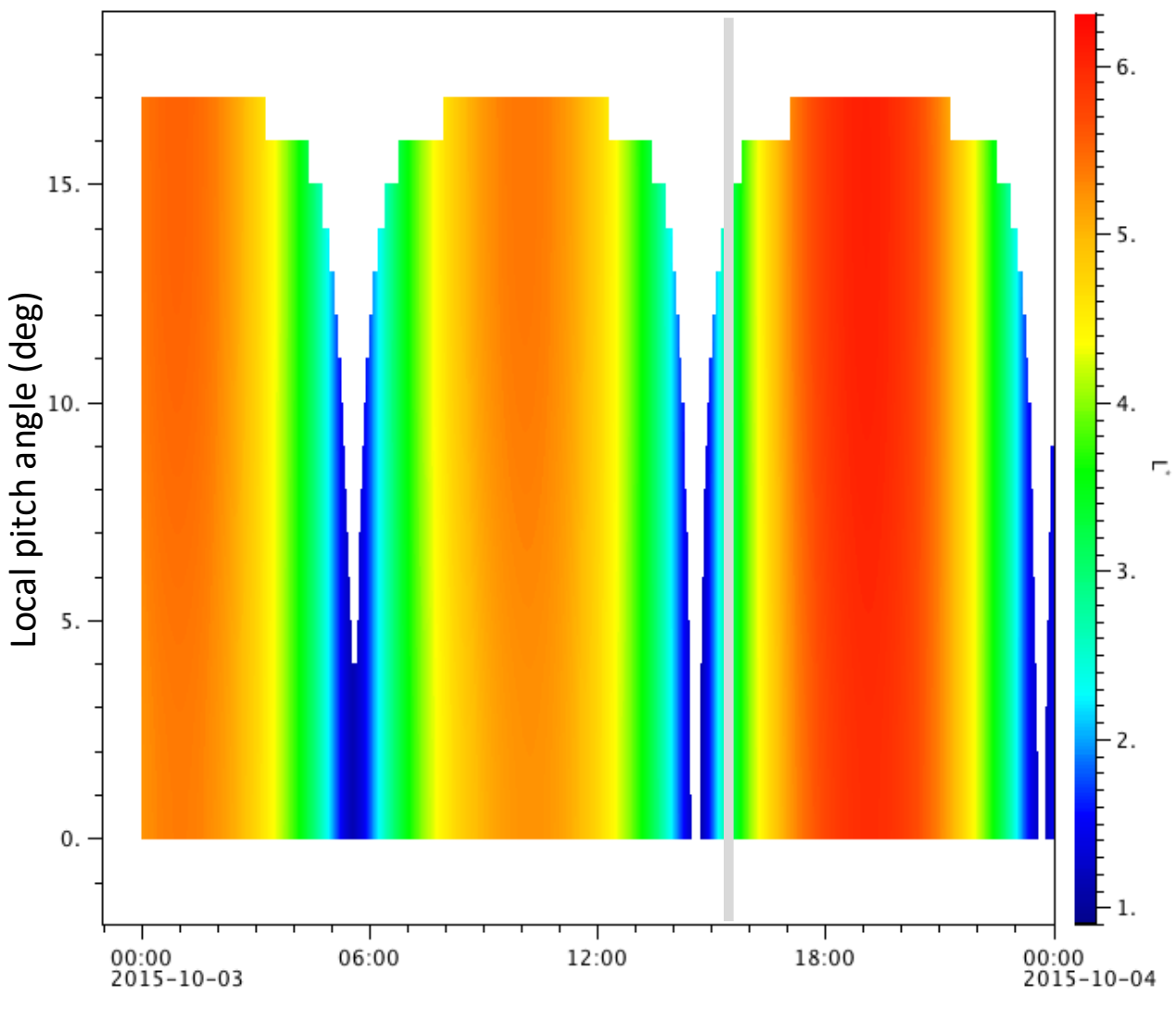

Fig. 11. Local pitch angle distribution from RBSP MagEIS. The gray line is the event time.

An important feature contributing to electron precipitation losses is the spread in their pitch angle distribution. Electrons with a wide range of equatorial pitch angles can bounce (mirror) from one conjugate hemisphere to the other as they drift between magnetic field lines. However, EMIC waves will modify the pitch angles such that more electrons find their pitch angles pushed towards the loss cone leading to atmospheric loss (precipitation) of the particle. Figure 11 shows the RBSP MagEIS local pitch angle distributions vs. time during this event. While measurements were not made coincident with the aircraft event (15 and 16

UT, gray line), measurements before and after the event show significant flux at low pitch angles near the loss cone at low L. This is further evidence for a plausible atmospheric loss mechanism for the relativistic electrons due to loss cone pitch angles.

While EMIC waves may be generated in the midnight sector of the magnetosphere during substorm particle injections, it is typical for wave growth to occur that results in transport of relativistic electrons to the dayside magnetosphere within minutes. The peak of the ARMAS measured event was at 11:38 local clock time, i.e., the aircraft observations were made in noon sector L-shells. Other sectors could have seen even more radiation. We conclude that EMIC waves were present within the time frame that the aircraft was observing elevated dose rates 03 October 2015.

**Fourth line of evidence (GEANT4 modeling corroborating results)**: Once the relativistic electrons hit the top of the atmosphere, which we consider to be 100 km, bremsstrahlung, photoelectric effect, Compton scattering, and pair production occur. From *electron* impact bremsstrahlung produces a γ-ray (1.24–124 MeV) or X-ray (0.124–1.24 MeV) (IS:21348, 2007) photon. Those high-energy *photon* can then induce one of the following processes: *i)* the photoelectric effect to produce an electron; *ii)* Compton scattering to eject a secondary inner shell (lower energy) electron plus a photon; or *iii)* pair production to eject a secondary inner shell electron plus a positron that then decays to an electron and photon. Thus, relativistic electron impacts on stratospheric $N_2$ and $O_2$, especially below 80 km, can cause production of lower energy secondary photons and electrons. The resultant spray of these secondaries and tertiaries can then impact other atmospheric target atoms causing additional lower energy photons and particles. Artamonov *et al.* (2016), using the CRAC:EPII model, have even demonstrated ion pair production in the lower atmosphere between 155–205 g cm$^{-2}$ (12–10 km) due to >10 MeV primary electrons producing secondary radiation. Here, we have used the SPENVIS GEANT4 (Agostinelli *et al.,* 2003; Allison *et al.,* 2006, 2016) model, which provides a useful platform for analyzing these processes with both electrons and photons.

We have conducted numerous GEANT4 model runs for neutrons, protons, electrons, positrons, and γ-rays using the geometrical assumption of injected charged particles from the radiation belts via a parallel beam influx or as secondary photons at the top of the atmosphere (Table 2) for the 03 October 2015 example. The version of GEANT4 that we use contains the Multi-Layered Shielding Simulation (MULASSIS) shielding code and is available on the SPENVIS website located at the URL https://www.spenvis.oma.be. MULASSIS is the relevant GEANT4 module for our study, instead of Planetocosmics, for example, since we want to obtain absorbed dose in silicon for comparison to ARMAS measurements. For the modeled run we created an atmosphere comprised of air (78% $N_2$ and 21% $O_2$ using the chemical formula of $N_4O$) from 100 km to 0 km (sea level). Table 2, using U.S. Standard Atmosphere densities to 20 km and MSIS-90 densities above 20 km to 100 km, shows air densities at each of the 17 altitude steps along with the modeled dose (rad) from bremsstrahlung γ-rays.

Table 2. Air altitude, density and dose

| Altitude (km) | Air density (mg cm$^{-3}$) | γ-ray dose (rad) |
|---|---|---|
| **100** | 5.36e-07 | 0.0000e+00 |
| **90** | 3.08e-06 | 1.4688e-05 |
| **80** | 1.44e-05 | 1.6285e-04 |
| **70** | 6.02e-05 | 3.0822e-04 |
| **60** | 2.39e-04 | 1.2362e-03 |
| **50** | 8.60e-04 | 3.2787e-03 |
| **40** | 3.26e-03 | 6.1879e-03 |
| **30** | 1.61e-02 | 5.4722e-03 |
| **20** | 8.57e-02 | 3.2009e-03 |
| **18** | 1.21e-01 | 1.4075e-03 |
| **16** | 1.65e-01 | 7.8268e-04 |
| **14** | 2.27e-01 | 3.1821e-04 |
| **12** | 3.11e-01 | 8.8355e-05 |
| **10** | 4.14e-01 | 1.9895e-05 |
| **08** | 5.25e-01 | 1.2509e-06 |
| **05** | 7.36e-01 | 0.0000e+00 |
| **00** | 1.23e+00 | 0.0000e+00 |

From our GEANT4 analysis we selected two cases to compare: *i)* exponentially fit electrons between 2 and 15 MeV shown in Figure 7 that are conservatively representative of the 03 October 2015 event conditions and *ii)* a 5 MeV mono-energetic γ-ray beam having an upper limit of 7% the fluence of the 2–15 MeV electrons. We demonstrate next that the most plausible explanation of what the aircraft experienced during the 03 October event was flight through a γ-ray beam; the electron modeling result is considered a least-case limit of possible dose.

To begin the comparison we first calculate the dose measured by ARMAS for the event. The aircraft was flying between L = 4.1 to L = 5.3 and measured absorbed dose (Si) at a constant altitude of 11.5 km. With the event start, ARMAS had already measured 4.620 μGy total dose from the secondary radiation field of protons and neutrons caused by GCRs; the GCR dose rate immediately prior to the event start time was 2.854 μGy $h^{-1}$. During the 33-minute event, another 2.10 μGy was measured and, if we subtract 1.57 μGy attributable to GCRs during this time frame (2.854×(33/60)), then the resulting difference of 0.53 μGy ($5.3\times10^{-5}$ rad) represents the amount of incremental dose by the energetic electron event that is not from GCRs or SEPs.

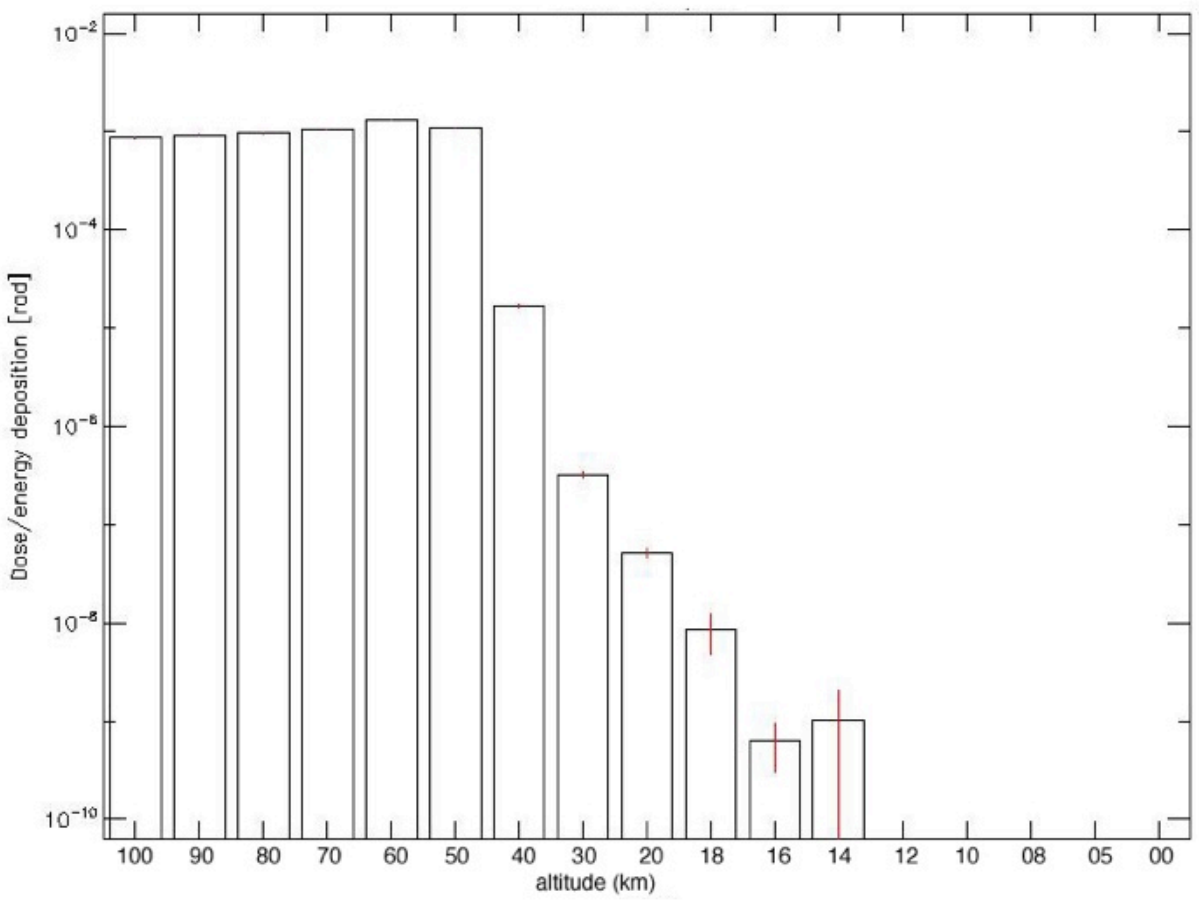

Fig. 12. GEANT4 2–15 MeV electrons exponentially fit in Figure 7 with a fluence of $6.43\times10^{7}$ electrons $cm^{-2}$ injected at the top of the atmosphere (Air) during a half-hour event using pair production as the electromagnetic cascade physics.

We next examine the GEANT4 MULASSIS model run for the relativistic electron precipitation (REP) case using the exponential fit to the primary electron energy spectrum derived from the RBSP-A REPT >2.3 MeV data (Figure 7, equation 1). A flux of $3.57\times10^{4}$ electrons $cm^{-2}$ $s^{-1}$ was calculated by GEANT4 using this spectrum. For the (rounded) 1800 seconds of the event, the fluence was $6.43\times10^{7}$ electrons $cm^{-2}$. By default, SPENVIS GEANT4 MULASSIS assumes pair production as the physics for the electromagnetic cascade that equally partitions the energy between electrons, positrons, and γ-rays with cutoff energies at 990 eV. Figure 12 shows that no dose is registered in the 12 km (aircraft) altitude bin using this physics although there is some dose with large uncertainty at the 14 km level. However, Hastings and Garrett (2004) identify that the photon energies in our range of interest (up to 15 MeV) impacting target atoms of low Z such as N (Z=7) and O (Z=8) have Compton scattering as the dominant energy transfer process and not pair production (Figure 13). If Compton scattering were available in SPENVIS GEANT4 so that energy was not wasted on positron production, theis produce higher dose at lower altitudes. The option to use this physics was not available in the online SPENVIS GEANT4 MULASSIS algorithm used for our runs. Thus, we take the electron modeling run as a least-case bounding condition. We also do not model EMIC-wave scattered electrons at higher energies from radial diffusion along with assuming an exponential fit is the best approximation to the actual (unknown) energy spectrum.

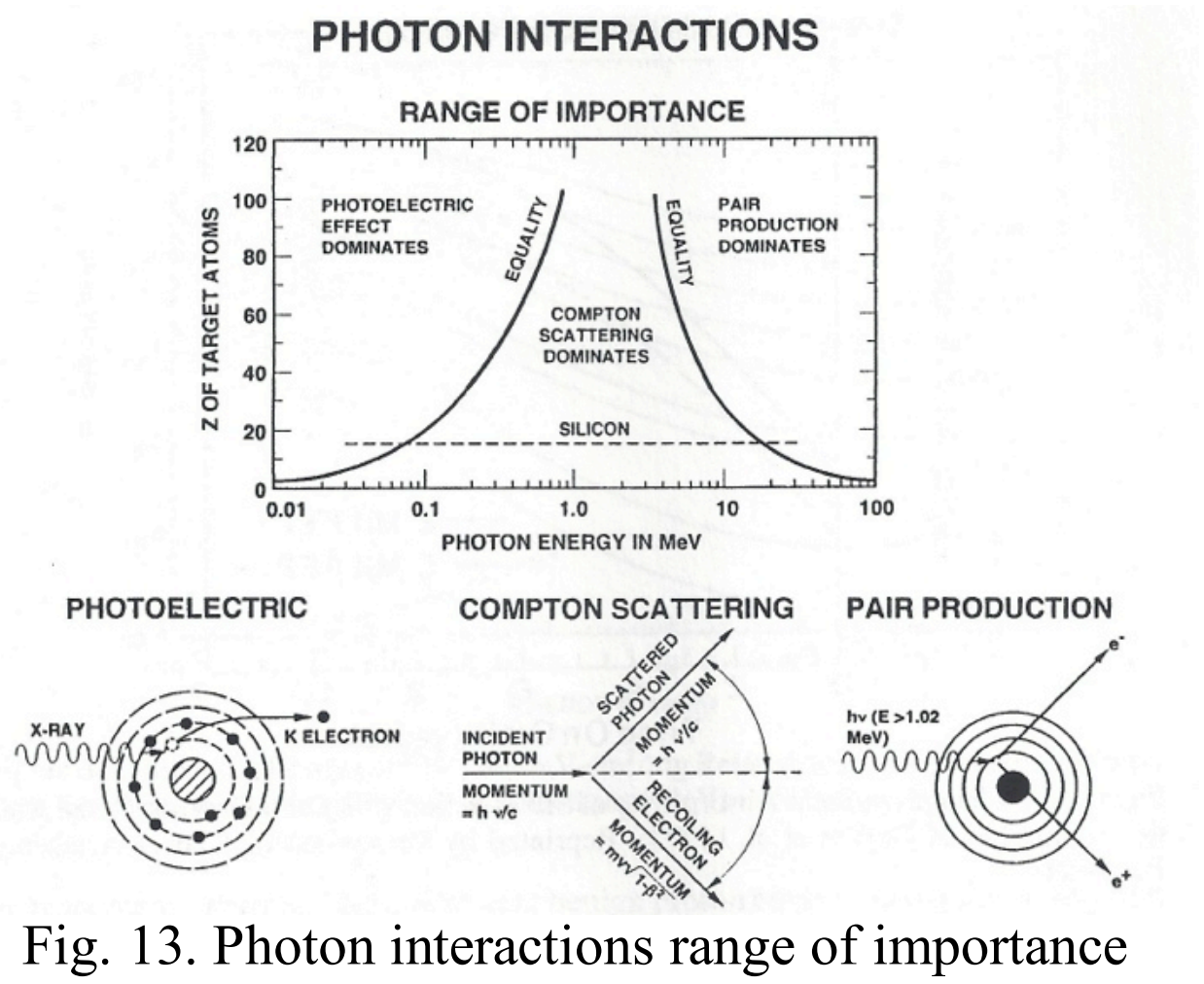

Fig. 13. Photon interactions range of importance showing dominating effects for photoelectric, Compton scattering, and pair production processes (from Hastings and Garrett, 2004).

Given this negative result using electrons, we then considered a second case. We first note the results of Carlson *et al.* (2008) who show that a seed population of 5–10 MeV electrons at the top of the atmosphere can produce secondary γ-ray photons with an efficiency of 10–50%. If we select a GEANT4 MULASSIS case with 5 MeV γ-ray photons in a parallel beam at the top of the atmosphere having 7% of the primary fluence (less than the Carlson *et al.* (2008) lower limit of 10%) then we obtain a fluence of $4.5 \times 10^6$ photons $cm^{-2}$ bremsstrahlung secondaries. The dose produced at 12 km is $8.8 \times 10^{-5}$ rad (Table 2) using these assumptions. The aircraft is 500 m below this altitude so a slightly lower dose would be observed. In fact, in this run 40% less dose of $5.3 \times 10^{-5}$ rad was measured by ARMAS (Figure 14 blue bar). This second model run is a bounding condition for the highest fluence of 5 MeV γ-rays needed to explain the measurements. Thus, this result plausibly suggests ARMAS is measuring dose resulting from aircraft flight through a bremsstrahlung-produced γ-ray beam.

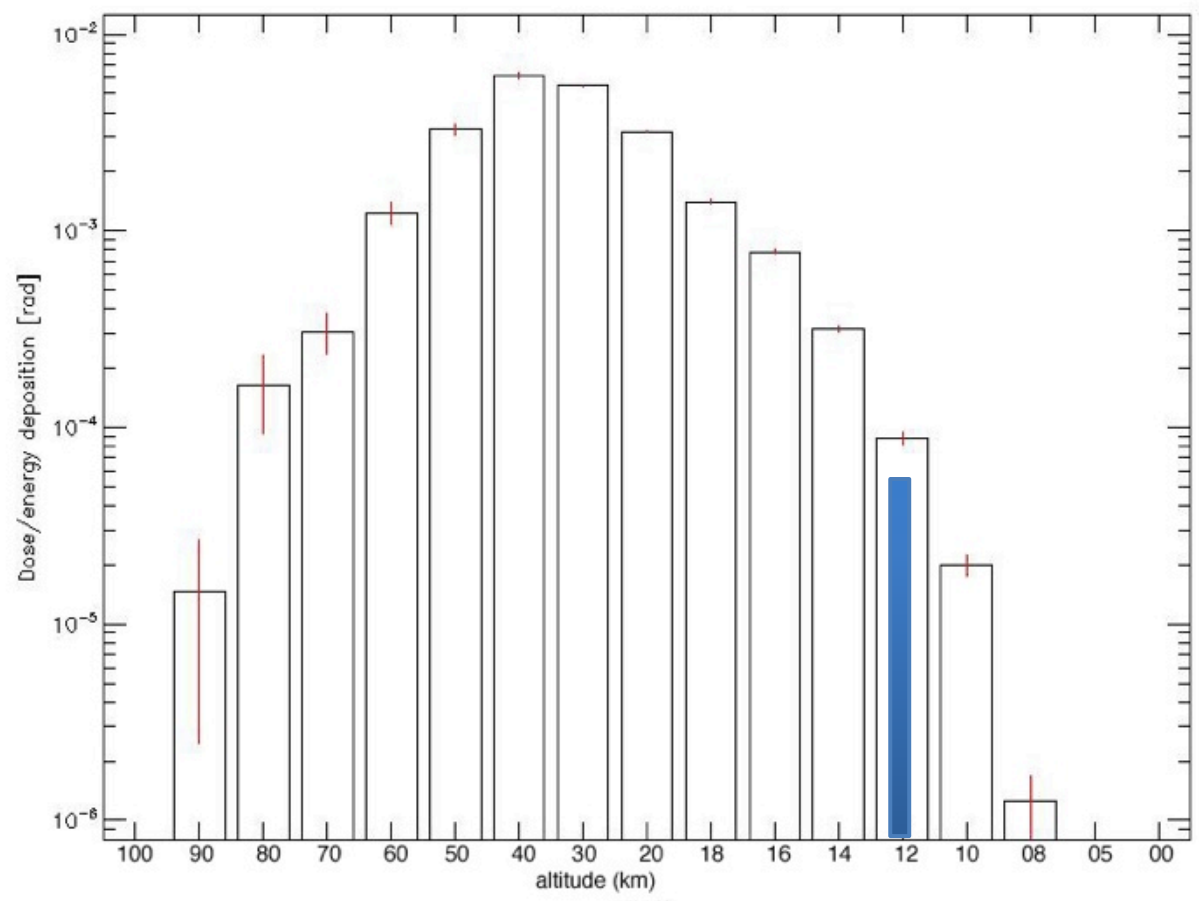

Fig. 14. GEANT4 mono-energetic 5 MeV γ-rays with a fluence of $4.5 \times 10^6$ photons $cm^{-2}$ injected at the top of the atmosphere (Air) during the half-hour event assumed from bremsstrahlung.

**03 October 2015 event results**: For the event the ARMAS FM2 instrument saw an enhanced dose rate at 11.5 km on 03 October 2015 for 33-minutes centered on 15:29 UT. The aircraft was on field lines mapping to 4.1 ≤ L ≤ 5.3 (Figure 15). We use the 2015 reference coordinates 80.4N and 72.6W for the North pole latitude and longitude to convert to magnetic coordinates and then obtain L values (Tascione, 1994). In Figure 15 the 03 October 2015 flight within the 11–12 km altitude layer (circled) is compared with all other ARMAS measurements between 2013 and 2019 during NOAA G-scale G0 conditions and across all L-shells. This flight is one of the few that stands out significantly (only atypical flights have been labeled with bold dots). We additionally show the D-index for tissue hazard (Meier and Matthiae, 2014; Tobiska *et al.*, 2017) with the color bar.

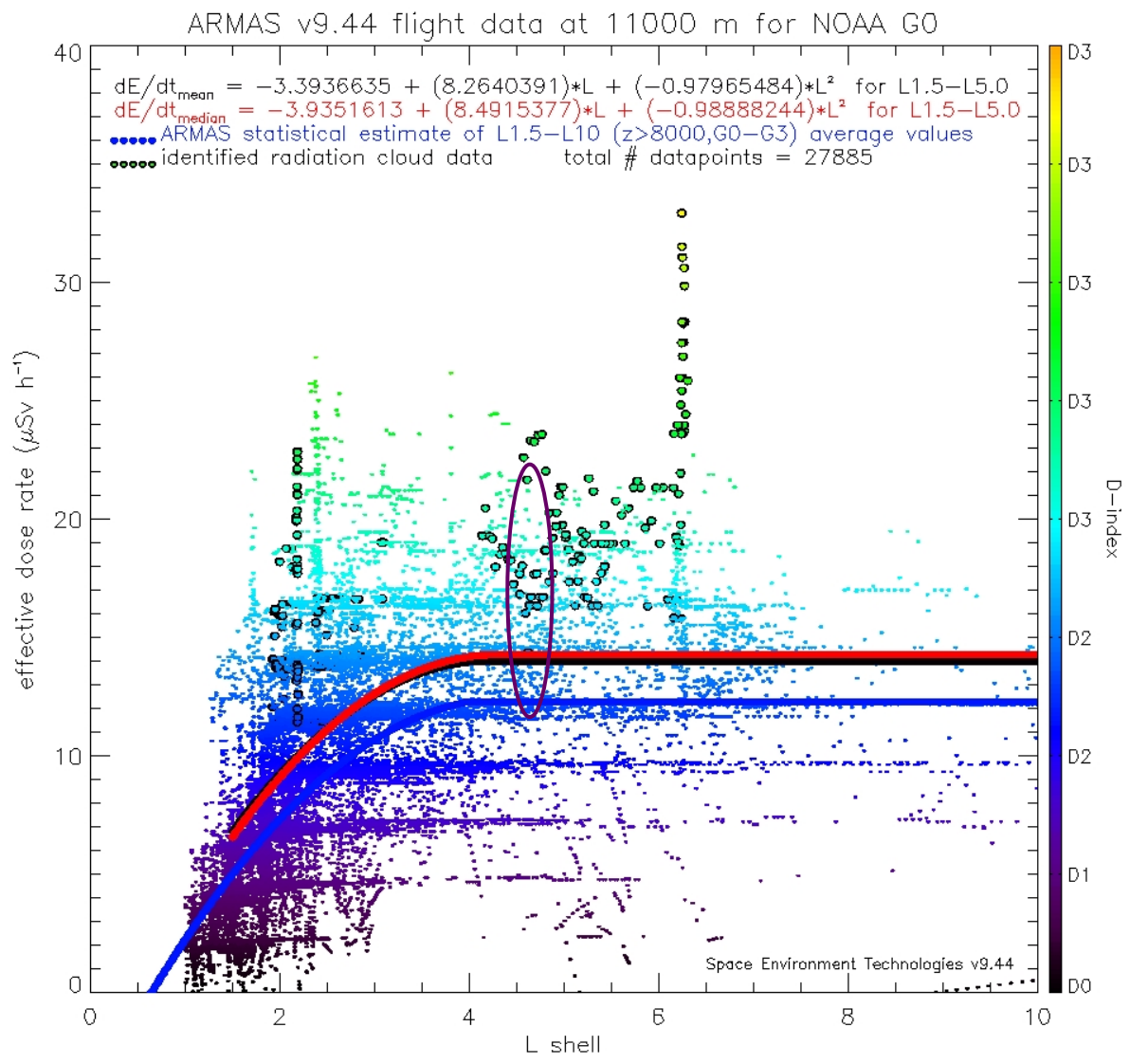

Fig. 15. ARMAS effective dose rates substantially increased between 4 ≤ L ≤ 5 during 03 Oct 2015 at 15–16 UT (circled) above typical 11 km GCR values for quiet geomagnetic conditions during solar cycle 24decline.

The heavy red and black lines through the data are the median and mean polynomial fits, respectively, that show the effect of radiation events combined with the GCR background compared to the heavy blue line of the GCR background only. The polynomial fit to the median GCR plus REP radiation data yields equation 2 for an effective dose rate, $\dot{E}$, in μSv h$^{-1}$ at 11 km as a function of L and during G0 conditions:

$$\dot{E}\,(L)_{11km,G0} = -0.78 + 7.38L - 0.78L^2 \quad (2)$$

where 1.5 ≤ L ≤ 5.0; the empirical representation, as a function of L-shell, constant altitude, and constant geomagnetic activity, of the slowly varying GCR background radiation component during 2013 – 2019, i.e., from solar cycle 24 decline to minimum is the black dotted line. There were no SEP sources of radiation during these flights. The ratio of the combined data to the GCR background alone at L-shell = 4 is 1.32 (16.3/12.3).

The observed dose (Si rad) of $5.3\times10^{-5}$ rad on the 03 October 2015 flight closely matches what a GEANT4 modeled dose would be with a fluence of $4.5\times10^{6}$ photons cm$^{-2}$ for 5 MeV γ-rays. The γ-rays reach their peak production at 40 km and are fully absorbed by 8 km based on GEANT4 modeling (Figure 14). This γ-ray fluence is consistent with using an upper limit of 7% of the primary relativistic electron fluence. The electrons were modeled with an exponentially fit energy spectrum between 2–15 MeV, having a fluence of $6.43\times10^{7}$ electrons cm$^{-2}$. The primary electrons begin to be absorbed near 50 km altitude (Figure 12) and, by 14 km, have been depleted. Figure 12 represents a least-case result using the default SPENVIS GEANT4 model with pair production physics for the electromagnetic cascade and equally partitioned energy between electrons, positrons, and photons.

From a modeling perspective it is possible we do not capture the fluence of other higher energy electrons created by radial diffusion or Chorus/EMIC wave acceleration when using an exponential fit approximation to the energy spectrum and, if this is the case, we could also potentially achieve the dose rates we see using Compton scattering physics. We also do not consider the shielding effects of different aircraft, which is an interesting topic but outside the scope of this study where we attempt to understand the external environment. We consider that γ-rays in the few MeV range and at fluence levels of less than 7% of the incident relativistic electrons are a viable explanation for the dose rates observed by ARMAS in this event.

**Conclusions.** First, we have plausibly identified a new source particle population that creates short-term, dynamic enhanced radiation phenomena at commercial aviation altitudes. Fifty-seven examples of enhanced radiation events have been found in our dataset consisting of 533655 one-minute measured absorbed dose (silicon) and derived effective dose rate records collected in 599 flights by the ARMAS aviation radiation monitoring program between 2013 and 2019.

The cloud concept is solely from the perspective of the aircraft. More accurately, it is more likely to be a flight through a bremsstrahlung produced γ-ray beam. Whether or not this event is temporal, spatial or a combination of both is still not fully determined. A beam formed above the aircraft would be consistent with an incident, precipitating flux of relativistic electrons along a parallel beam.

We explored one case of the fifty-seven examples in detail. The electron beam was aligned with magnetic field lines having L-shells between 4.1 and 5.3, which map into the outer Van Allen radiation belt. In this case (#13 in Table 1), we have identified processes whereby the 2–15 MeV relativistic electrons can be lost to the atmosphere. This occurs because their pitch angles are scattered into the loss cone at >15° local angles by EMIC waves. These EMIC waves were generated in the outer radiation belt by geomagnetic substorm particle injections and the disturbed electrons quickly made their way into the magnetosphere noon sector.

Considering all fifty-seven examples, the event durations from tens of minutes to over an hour suggest a temporal characteristic for these radiation event. However, the width of a secondary and tertiary radiation field caused by bremsstrahlung from the precipitating primary electrons has a distance that is at least as wide as the range of magnetic fields lines along which electrons are precipitating. Thus, we find it is reasonable to consider that the radiation field may be spatial as well. Of the fifty-seven examples listed in Table 1 we note that events range from 15 – 99 minutes in duration. At normal aircraft velocities in the upper troposphere of 600 – 700 km $h^{-1}$, the spatial extent of these events ranges from 150 – 1150 km. We note that the spatial extent of these events is consistent with relativistic electron precipitation bands described by Blake *et al.* (1996) and observed outside the International Space Station (ISS) by Dachev *et al.* (2017) and Dachev (2017). Table 1 also identifies numerous examples of events at L-shells that would map to the inner Van Allen radiation belts; however, we have not explored those events in detail.

Second, the mechanisms that lead to an increased exposure hazard at commercial aviation altitudes >8 km and higher magnetic latitudes occur frequently even under minor geomagnetic substorm conditions. The magnetic latitudes we observed to be most affected are those between 43° – 67° in both hemispheres. In the northern hemisphere these are the latitudes used by the North Atlantic Traffic (NAT), the North Pacific (NoPAC) and the northern half of CONUS. This suggests a higher statistical career exposure to aircrew members and frequent flyers on NAT, NoPAC, and northern CONUS routes than can be estimated from GCR radiation alone. In fact, the ratio of all ARMAS 11–12 km flights for NOAA G0 conditions compared to GCR background from NAIRAS is 1.32. The mean of all the Table 1 ARMAS flight effective dose rates, for the duration of all the flights, as a ratio to the NAIRAS GCR dose rates, is 1.43 and, for the events themselves, there is a doubling (1.91) of the exposure compared to GCR background.

Third, in Figure 15, we have identified an average GCR background effective dose rate background at 11 km and under constant NOAA G-scale G0 geomagnetic activity as a function of L-shell during the 2013 – 2019 solar cycle 24 decline to minimum. When we include GCRs and REP enhanced radiation events we get an effective dose rate as a function of L-shell (equation 2). We have created a median statistical database for most locations around the planet at altitudes from 8 to 19 km in 1 km steps, at $1.5 \leq L \leq 6.0$, and for G0 to G4 geomagnetic conditions. An excerpt from that database is shown in Figure 15; the full results will be presented in a subsequent paper. Effective dose rate values of 9 μSv $h^{-1}$ for low L locations (L = 1.5 for tropical and equatorial latitudes) during G0 quiet geomagnetic conditions at 11 km and values of 16–17 μSv $h^{-1}$ for high L locations (L > 4 for northern and southern high mid-latitudes, not polar) are consistent with NAIRAS GCR background effective dose rates, as shown in Figure 2 (top panel), plus a REP radiation event component.

Other mechanisms not considered, but possible for creating this enhanced radiation include HISS, whistler, chorus, plasmasheet curvature scattering. Also using L-shell instead of L*.

**Acknowledgments:** The authors thank the reviewers for their timely and insightful comments that have improved this paper. The authors acknowledge the financial support for ARMAS from the original NASA NAIRAS DECISION project contract NNL07AA00C, the NASA SBIR Phase I and Phase II program contracts NNX11CH03P and NNX12CA78C, the NASA AFRC Phase III contracts NND14SA64P and NND15SA55C, the NASA LWS TRT RADIAN grant 80NSSC18K0187, and the South Korean Space Weather Center matching funds for SBIR Phase IIE. The NASA Airborne Sciences Program and Armstrong Flight Research Center DC-8 and Gulfstream 3 have provided welcome flight support for ARMAS instruments. E. Teets, S. Wiley, T. Moes and R. Albertson have enabled successful measurements from AFRC while L. Guhathakurta of NASA programmatically contributed to the G-3 success. The NOAA Space Weather Prediction Center used their good offices to facilitate ARMAS use on NOAA Gulfstream 4 as did the National Center for Atmospheric Research High Altitude Observatory for ARMAS use on the National Science Foundation Gulfstream 5. The University of Alaska,

Fairbanks Geophysical Institute also provided welcome collaboration for obtaining ARMAS FM5 flight data on commercial flights at higher latitudes across North America. The Korean Space Weather Center provided useful collaboration for obtaining ARMAS FM5 flight data on commercial flights at higher latitudes across the North Pacific. The ARMAS Level 4 archival data are publicly available from the ARMAS website at the URL http://sol.spacenvironment.net/~armas_ops/Archive/ and the NAIRAS real-time data is at http://sol.spacenvironment.net/raps_ops/current_files/globeView.html.